\documentclass[12pt, reqno]{amsart}
\usepackage{amssymb}
\usepackage{mathrsfs}
\usepackage{mathpazo}
\usepackage[colorlinks=true, breaklinks=true]{hyperref}
\usepackage{bm}
\usepackage[latin1]{inputenc}
\usepackage{graphicx}
\newtheorem{theorem}{Theorem}

\newtheorem{definition}[theorem]{Definition}

\newtheorem{lemma}[theorem]{Lemma}

\newtheorem{proposition}[theorem]{Proposition}
\newtheorem{remark}[theorem]{Remark}

\newcommand{\bea}{\begin{eqnarray}}
\newcommand{\eq}{\end{eqnarray}}
\newcommand{\eea}{\end{eqnarray}}
\newcommand{\bqn}{\begin{eqnarray*}}
\newcommand{\beaa}{\begin{eqnarray*}}
\newcommand{\eqn}{\end{eqnarray*}}
\newcommand{\eeaa}{\end{eqnarray*}}
\newcommand{\bpr}{\begin{proposition}}
\newcommand{\epr}{\end{proposition}}
\newcommand{\nn}{\nonumber}
\newcommand{\cal}{\mathcal}
\begin{document}

\def\BS{\rm BS}

\begin{center}
 \large \bf{The Heston Riemannian Distance Function}
\end{center}
\vspace{1cm}
\begin{center}
\bf Archil Gulisashvili\,$^{a,*}$,\quad Peter Laurence\,$^b$ \rm
\end{center}
\vspace{0.3in}
\footnotesize\it $^a$\,Department of Mathematics, Ohio University, Athens, OH 45701, USA
\\
\\
$^b$\,Dipartimento di Matematica, Universit\`{a} di Roma 1,
Piazzale Aldo Moro, 2, I-00185 Roma, Italia, and
Courant Institute of Mathematical Sciences, New York University,
251 Mercer Street, New York, NY 10012, USA \normalsize\rm
\\
\begin{center} 
Abstract
\end{center}
\vspace{0.1in}
The Heston model is a popular stock price model with stochastic volatility that has found numerous applications in practice. In the present paper, we study the Riemannian distance function associated with
the Heston model and obtain explicit formulas for this function using geometrical and analytical methods. Geometrical approach is based on the study of the Heston geodesics, while the analytical approach
exploits the links between the
Heston distance function and the sub-Riemannian distance function in the Grushin plane. For the Grushin plane, we establish an explicit formula for the Legendre-Fenchel transform of the limiting cumulant generating function and prove
a partial large deviation principle that is true only inside a special set.  
\vspace{0.1in}
\section{Introduction}\label{S:HeGr}
There are two main protagonists
 in this paper: the Riemannian manifold associated with the Heston model of the stock price, and the Grushin plane, which is one of the best-known examples of a sub-Riemannian manifold. The present paper focuses on the Heston Riemannian distance and the Grushin sub-Riemannian distance and provides explicit formulas for these. The Heston distance and the Grushin distance are intimately related, and various facts concerning these distances can be easily transplanted from one  setting into the other.

We will next briefly describe the main results obtained in this paper. Theorems \ref{T:Hd1} and \ref{T:Hd2} below contain explicit formulas for the Heston distance. The formulas in Theorem \ref{T:Hd1} are established using geometrical methods, while the proof of the distance formula in Theorem \ref{T:Hd2} uses certain links between the Heston and the Grushin distances and is more analytical. In the proof of Theorem
\ref{T:Hd2}, we compute and study the limiting cumulant generating function $\Lambda$ for the Grushin plane and the Legendre-Fenchel transform $\Lambda^{*}$ of the function $\Lambda$. One of the main results in the present paper is a partial large deviation principle for the Grushin plane (see Theorem \ref{T:sum}). The word ``partial" is used in the previous sentence because in the case of the Grushin plane the large deviation principle with $\Lambda^{*}$ as a rate function holds only inside a special subset
of $\mathbb{R}^2\times\mathbb{R}^2$. We would also like to bring the reader's attention to the results concerning certain qualitative properties of the transcendental equations whose solution is involved in determining the Heston Riemannian distance function. These convexity and monotonicity
properties established in Lemmas \ref{L:hidden}, \ref{L:nohom}, and \ref{L:cancan0} ensure that the equations can be efficiently and rapidly solved
by Newton's method or a bisection method. We also show in the present paper that it is crucial to distinguish two different regimes (the near and the far point regime) in the geometrical and analytical approaches to the Heston distance, each regime requiring it's
own analysis (see Theorems \ref{T:Hd1} and \ref{T:Hd2} and their proofs).

Let us expand on the financial motivations for considering the Heston Riemannian distance function.
The Heston model is one of the most popular stock price models with stochastic volatility. This model was introduced in \cite{H}. More information on the Heston model and stochastic volatility models can be found
in \cite{FPSS,G,H-L,ZS}. The stock price process $S$ and the variance process $V$ in the Heston model satisfy the following system of stochastic differential equations:
\begin{equation}
\left\{\begin{array}{ll}
dS_t=\mu S_tdt+\sqrt{V_t}S_tdW_t \\
dV_t=(a-bV_t)dt+c\sqrt{V_t}dZ_t,
\end{array}
\right.
\label{E:Heston1}
\end{equation}
where $a\ge 0$, $b\ge 0$, $c> 0$. In (\ref{E:Heston1}), $W$ and $Z$ are correlated standard Brownian motions
such that $d\langle W,Z\rangle_t=\rho dt$ with $\rho\in(-1,1)$.

Recently closely related models, the local-volatility Heston models,
given by
$dS_t=\mu S_tdt+\sqrt{V_t} \sigma(S_t) S_tdW_t$
for an appropriate function $\sigma$,  have become objects of widespread interest
among practitioners. Practitioners seek to come up with accurate approximations to the Black-Scholes
implied volatility in such models and there is a considerable literature in this
direction. In one of the approaches to this problem, initiated for another
class of stochastic volatility models in \cite{HagLes}, and in the Heston case
by \cite{FJ1}, \cite{ForJacMij}, \cite{ForJacLee} a key element in determining the implied volatility is
the Riemannian distance to a line $S=K$ in the $SV$-plane.
We plan to address applications of the results obtained in the present paper to the local volatility Heston models in future publications. This will include consideration of
heat kernel expansions, implied volatility expansions, and pricing of exotic options in such models.

Let us consider the following uncorrelated Heston model:
\begin{equation}
\left\{\begin{array}{ll}
dS_t=S_t\sqrt{V_t}dW_t \\
dV_t=(a-bV_t)dt+\sqrt{V_t}dZ_t,
\end{array}
\right.
\label{E:Heston11}
\end{equation}
where $a\ge 0$, $b\ge 0$, and $W$ and $Z$ are independent standard Brownian motions.
Denote by $X$ the log-price process defined by $X=\log S$. Then the model in (\ref{E:Heston11})
transforms as follows:
\begin{equation}
\left\{\begin{array}{ll}
dX_t=-\frac{1}{2}V_tdt+\sqrt{V_t}dW_t \\
dV_t=(a-bV_t)dt+\sqrt{V_t}dZ_t.
\end{array}
\right.
\label{E:Hes-ton1}
\end{equation}
The state space for the process $(X,V)$ is the closed half-plane
$$
{\cal H}=\left\{(x,v)\in\mathbb{R}^2:v\ge 0\right\}.
$$
We will denote the initial condition for the process $(X,V)$ by $(x_0,v_0)$.

The Riemannian metric form associated with the Heston model is defined on the interior ${\cal H}^{\circ}$
of the closed half-plane ${\cal H}$ as follows:
\begin{equation}
ds^2=v^{-1}\left(dx^2+dv^2\right).
\label{E:form1}
\end{equation}
The open half-plane
${\cal H}^{\circ}$ equipped with the metric defined in (\ref{E:form1}) is called the Heston manifold. The form in (\ref{E:form1}) generates the Riemannian distance $d_H$ on ${\cal H}$. More examples of Riemannian
distances arising in finance can be found in \cite{H-L} (see also \cite{L}).

In this paper, we discuss various explicit formulas for the Heston distance
$d_H$. It is worth mentioning that the following two-sided estimate for $d_H$ is known
(see \cite{K}, Proposition 4.3.2):
\begin{equation}
D(x_0,v_0,x_1,v_1)\le d_H((x_0,v_0),(x_1,v_1))\le 12D(x_0,v_0,x_1,v_1)
\label{E:twos1}
\end{equation}
for all $(x_0,v_0)\in{\cal H}$ and $(x_1,v_1)\in{\cal H}$,
where
\begin{equation}
D(x_0,v_0,x_1,v_1)=\frac{\sqrt{(x_0-x_1)^2+(v_0-v_1)^2}}{\sqrt{v_0}+\sqrt{v_1}+[(x_0-x_1)^2+(v_0-v_1)^2]^{\frac{1}{4}}}.
\label{E:twos3}
\end{equation}

We will next briefly explain how to obtain explicit formulas for the distance function
$\widetilde{d}_H$ in the general correlated Heston model described in
(\ref{E:Heston1}) from similar formulas for the distance function $d_H$ in the uncorrelated Heston model with the vol of vol coefficient equal to one considered in the present paper.
It is known that the principal part of the generator of the diffusion $(X,V)$ with $X=\log S$ in model
(\ref{E:Heston1}) is given by
\beaa
v\left(\frac{\partial^2}{\partial x^2} + 2 \rho c\frac{\partial^2}{\partial x\partial v} + c^2
\frac{\partial^2}{\partial v^2}\right).
\eeaa
Let us show how to reduce this to the case of an uncorrelated Heston model, in which the metric
is in the standard form presented in (\ref{E:form1}). First make the change of time
$\hat{t}= c^2 t$ which reduces the principal part of the generator of the diffusion
to
\beaa
v\left(c^{-2}\frac{\partial^2}{\partial x^2} + 2 \rho c^{-1}\frac{\partial^2}{\partial x\partial v} +
\frac{\partial^2}{\partial v^2}\right).
\eeaa
It is not hard to see that under such a change of time, the distance function $\widetilde{d}_H$ is multiplied by the constant $c$. Next, we use the diffeomorphism
\beaa
&& \hat{x} = \frac{c}{\sqrt{1 - \rho^2}} x - \frac{\rho}{\sqrt{1 - \rho^2}} v
\\
&& \hat{v} = v.
\eeaa
After this change of variables the principal part of the new diffusion operator
is in the standard form
\beaa
\hat{v}\left(\frac{\partial^2}{\partial\hat{x}^2}+\frac{\partial^2}{\partial\hat{v}^2}\right).
\eeaa
Taking into account the reasoning above, one can prove that
\begin{align}
&\widetilde{d}_H((x_0,v_0),(x_1,v_1)) \nonumber \\
&=\frac{1}{c}d_H\left(\left(\frac{cx_0-\rho v_0}{\sqrt{1-\rho^2}},v_0\right),\left(\frac{cx_1-\rho v_1}{\sqrt{1-\rho^2}},v_1\right)\right).
\label{E:sveli}
\end{align}
Formula (\ref{E:sveli}) shows how to adapt the distance formulas obtained in this paper to the case
where the Heston model is correlated and given by (\ref{E:Heston1}).

Let us note that the drift terms in (\ref{E:Hes-ton1})
do not affect the Heston distance. On the other hand, the Heston transition density
$(x,v)\mapsto p_t^H((x_0,v_0),(x,v))$
associated with the process $(X,V)$ is influenced by the drift terms.
Using the definition of the Riemannian distance and (\ref{E:form1}), we see that the distance $d_H$ satisfies the following conditions:
\begin{equation}
d_H((x_0,v_0),(x_1,v_1))=d_H((0,v_0),(x_1-x_0,v_1))
\label{E:form2}
\end{equation}
and
\begin{equation}
d_H((\alpha x_0,\alpha v_0),(\alpha x_1,\alpha v_1))=\sqrt{\alpha}
d_H((x_0,v_0),(x_1,v_1))
\label{E:form3}
\end{equation}
for all $\alpha> 0$.

It will be assumed throughout the paper that $x_0\neq x_1$. In the case where $x_0=x_1$,
the geodesics joining the points $(x_0,v_0)$ and $(x_1,v_1)$ is a vertical line
with length
\beaa
\int_{\min(v_0,v_1)}^{\max(v_0,v_1)} \frac{1}{\sqrt{v}} dv = 2 (\sqrt{\max(v_0,v_1)}
- \sqrt{\min(v_0,v_1)}),
\eeaa
and hence
$$
d_H((x_0,v_0),(x_0,v_1))=2 (\sqrt{\max(v_0,v_1)}
- \sqrt{\min(v_0,v_1)}).
$$
It follows from the previous formula that the $x$-axis is at a {\it finite} distance from any point in ${\cal H}^{\circ}$, and hence
the $x$-axis, being part
of the boundary of the Heston manifold, is "at infinity".
Therefore, the Heston manifold is {\it not complete},
and we can not apply the Hopf-Rinow theorem to establish the existence of a length-minimizing geodesic joining two points in ${\cal H}$. Note that one difficulty
in establishing such a result is proving that the length minimizing curve
joining {\it any} two points is a true geodesic and not the
union of broken geodesics (on the other hand for the metric $ds^2 = v( d x^2 + d y^2)$, Robert Bryant
has communicated to us that only a subset of points in the upper half space can be joined by a non-broken geodesic). 

It is interesting that for the metric defined by (\ref{E:form1}) the existence and uniqueness result for the length-minimizing geodesics has been essentially known for at least one century. We will next provide more information.

Let us consider the following metric form:
$$
ds^2=(2v)^{-1}\left(dx^2+dv^2\right)
$$
that is intimately related to the Heston metric. It is known that the length-minimizing geodesics for this metric are dilations and shifts of the
standard cycloid given by
\begin{equation}
\left\{\begin{array}{ll}
x=s-\sin s \\
v=1-\cos s.
\end{array}
\right.
\label{E:scyc}
\end{equation}
This was established by O. Bolza in 1904 (see \cite{B1}, see also \cite{HD}, Proposition I.2.1). Bolza proved that for any two points $C=(x_0,v_0)$ and $D=(x_1,v_1)$ in ${\cal H}$ with $x_0\neq x_1$ there exists exacty one dilated and shifted standard cycloid joining $C$ and $D$ and such that no cusp of the cycloid lies between $C$ and $D$. Bolza states that the previous result was included without proof in unpublished lectures of K. T. W. Weierstrass (1882). A special case was handled by Johann and Jacob Bernoulli at the very end of the 17th century.

It follows from the above-mentioned result of Bolza
that the Heston geodesics
can be obtained from the cycloid curve
\begin{equation}
\left\{\begin{array}{ll}
x=\frac{1}{\sqrt{2}}s-\sin\frac{s}{\sqrt{2}} \\
v=1-\cos\frac{s}{\sqrt{2}}
\end{array}
\right.
\label{E:cycloid}
\end{equation}
by translation $x\mapsto x+b$ and dilation $x\mapsto cx$, $v\mapsto cv$, or are vertical lines. It is also clear that Bolza's existence and uniqueness theorem holds for the Heston manifold.
\section{The Heston distance}\label{S:HDGA}
For every fixed $C\in\mathbb{R}$ such that $C\neq 0$, define a function on the interval
$[0,C^{-2}]$ by
\begin{equation}
\overline{F}(v,C)=\frac{\arcsin(C \sqrt{v})}{C^2}-\frac{\sqrt{v} \sqrt{1-C^2 v}}{C}.
\label{E:function1}
\end{equation}
Now fix $v_0\ge 0$ and $v_1\ge 0$ and consider a function of the variable $C$ given by
\begin{equation}
F(v_0, v_1, C)=\overline{F}(v_1,C)-\overline{F}(v_0,C).
\label{E:C-funct}
\end{equation}
It is assumed in (\ref{E:C-funct}) that the variable $C$ satisfies
the condition
$$
0<|C|\le\min\left(v_0^{-\frac{1}{2}},v_1^{-\frac{1}{2}}\right).
$$
Since $\displaystyle{\lim_{C\rightarrow 0}F(v_0, v_1, C)=0}$ (use l'H$\rm\hat{o}$pital's rule twice),
we can extend the function $F$ continuously to an odd function on the interval
\begin{equation}
I=[-\min(v_0^{-\frac{1}{2}},v_1^{-\frac{1}{2}}),
\min(v_0^{-\frac{1}{2}},v_1^{-\frac{1}{2}})]
\label{E:interv}
\end{equation}
by putting $F(v_0,v_1,0)=0$.

We will next prove a theorem that provides explicit formulas for the Heston distance. Note that there are two expressions for the Heston distance in Theorem \ref{T:Hd1},
depending on the location of the points $(x_0,v_0)$ and $(x_1,v_1)$ in the Heston half-plane.
\begin{theorem}\label{T:Hd1}
(i)\,\,Suppose the points $(x_0,v_0)$ and $(x_1,v_1)$ in the Heston half-plane satisfy the following condition:
\begin{equation}
|x_1 - x_0| \leq  F(min(v_0, v_1), max(v_0, v_1), \frac{1}{\sqrt{max(v_0, v_1)}}),
\label{E:c-reg1}
\end{equation}
where $F$ is given by (\ref{E:C-funct}). Then
\begin{align}
&d_H((x_0,v_0),(x_1,v_1)) \nonumber \\
&=2\frac{\arcsin(C^* \sqrt{max(v_0,v_1)})
-\arcsin(C^* \sqrt{min(v_0,v_1)})}{ C^*},
\label{E:nes1}
\end{align}
where $C^{*}=C^{*}((x_0,v_0),(x_1,v_1))$ is the unique solution to the transcendental equation
\begin{align}
x_1-x_0&=\frac{\sqrt{v_0} \sqrt{1-C^2 v_0}}{C}-\frac{\arcsin(C \sqrt{v_0})}{C^2}
\nonumber \\
&\quad-\left[\frac{\sqrt{v_1} \sqrt{1-C^2 v_1}}{C}-\frac{\arcsin(C \sqrt{v_1})}{C^2}\right]
\label{E:nes2}
\end{align}
on the interval $I$ defined by (\ref{E:interv}).
\\
\\
(ii)\,\,Suppose the points $(x_0,v_0)$ and $(x_1,v_1)$ in the Heston half-plane
satisfy the following condition:
\begin{equation}
|x_1 - x_0| \ge  F(min(v_0, v_1), max(v_0, v_1), \frac{1}{\sqrt{max(v_0, v_1)}}).
\label{E:c-reg2}
\end{equation}
Then
\begin{equation}
d_H((x_0, v_0), (x_1, v_1))=2\frac{\arccos(C^* \sqrt{v}_0) +\arccos(C^* \sqrt{v}_1))}{C^*},
\label{E:nes3}
\end{equation}
where $C^{*}=C^{*}((x_0,v_0),(x_1,v_1))$ is the unique solution to the
transcendental equation
\begin{align}
|x_1 - x_0|&=\frac{\sqrt{ v_0}\sqrt{1-C^2 v_0}}{C}+\frac{\arccos(C \sqrt{v_0})}{C^2}
\nonumber \\
&\quad+ \frac{\sqrt{v_1} \sqrt{1-C^2 v_1}}{C}+\frac{\arccos(C \sqrt{v_1})}{C^2}
\label{E:nes4}
\end{align}
on the interval $[0,\min(v_0^{-\frac{1}{2}},v_1^{-\frac{1}{2}})]$.
\end{theorem}

\it Proof. \rm We will first prove part (i). It is not hard to see that with no loss of generality
we may assume that $x_0< x_1$ and $v_0< v_1$. Let $c$ be the dilation coefficient corresponding to the points $A=(x_0,v_0)$ and
$B=(x_1,v_1)$ in Bolza's description of the geodesics. Then we have $c> 0$.
Define $C> 0$ from the equality $c=2C^2$, and consider the geodesic, connecting $A$ with $B$, under the
scaled parametrization $s\mapsto Cs$.

Let us first assume that the point $A$ and $B$ are both to the left of
the apex of the arc of the geodesic passing through them. Then it is not hard
to see that the components
$s\mapsto x(s)$ and $s\mapsto v(s)$ of the arc of the geodesic through $A$ and
$B$ satisfy
\begin{equation}
\left\{\begin{array}{ll}
\dot{x}= C v \\
\dot{v}= \sqrt{v} \sqrt{ 1 - C^2 v},
\end{array}
\right.
\label{E:cycloids}
\end{equation}
where the derivative in the system above is taken with respect
to the arclength parameter and where $v\in[0,C^{-2}]$. It is also clear
that $x$ is a function of $v$ for $v\in[0,C^{-2}]$ and we have
\bea
\frac{d x}{d v} = \frac{C \sqrt{v}}{\sqrt{1 - C^2 v}}.
\label{Csolution1}
\eea
Note that $v=C^{-2}$ is the second component of the apex. Therefore $[v_0,v_1]\subset[0,C^{-2}]$.

It is easy to see that
\begin{equation}
\frac{d\overline{F}}{d v} = \frac{C \sqrt{v}}{\sqrt{1 - C^2 v}}.
\label{E:Csol}
\end{equation}
It follows from (\ref{Csolution1}) and (\ref{E:Csol}) that
$\overline{F}(v,C)=x(v)+\alpha$,
where $\alpha$ is some constant. Plugging $v=0$ into the previous equality, we get
$\alpha=b$, where $b$ is the shift parameter in the description of the geodesic
passing through $A$ and $B$. Hence,
\begin{equation}
\overline{F}(v,C)=x(v)+b.
\label{E:Csol1}
\end{equation}
It follows from (\ref{E:Csol1}) that $C$ satisfies the following condition:
$$
x_1-x_0=\overline{F}(v_1,C)-\overline{F}(v_0,C)=F(v_0,v_1,C).
$$
Moreover, the assumptions formulated above can be formulated as follows:
$$
x_1-x_0\le F\left(v_0,v_1,\frac{1}{\sqrt{v_1}}\right).
$$

It is not hard to see using (\ref{Csolution1}) that
\begin{align*}
d_H(A,B)&=\int_{v_0}^{v_1}
\sqrt{\left(\frac{d x}{dv}\right)^2 + 1}\frac{dv}{\sqrt{v}}
=\int_{v_0}^{v_1}
\frac{1}{\sqrt{1-C^2v}}\frac{dv}{\sqrt{v}} \\
&=2\frac{\arcsin(C\sqrt{v_1})
-\arccos(C\sqrt{v_0})}{C}.
\end{align*}
This establishes formula (\ref{E:nes1}), and completes the proof of part (i) of Theorem
\ref{T:Hd1} in the case where the points $A$ and $B$ are both to the left of
the apex of the arc of the geodesic passing through them. The proof of part (i) in the case where
$A$ and $B$ are to the right of the apex is similar.

Next suppose that one of the points is to the left of the apex, while the other one is to its right.
This case is a combination of the previous two.
With no loss of generality, we may assume that $(x_0, v_0)$ is
to the left of the apex and $(x_1, v_1)$ is to the right of the apex. This happens if and only if
condition (\ref{E:c-reg2}) holds.
It is not hard to see that under the restriction imposed above, we need to sum two contributions, one going from $(x_0, v_0)$ to the apex and
the other going from the apex to $(x_1, v_1)$.
Since the $v$-component of the apex equals $C^{-2}$, we obtain
\begin{align*}
x_{\mbox{\tiny apex}} - x_0&= \frac{\sqrt{v_0} \sqrt{1-C^2 v_0}}{C} -\frac{\arcsin(C \sqrt{v_0})}{C^2}+ \frac{\pi}{2 C^2} \\
&=\frac{\sqrt{v_0} \sqrt{1-C^2 v_0}}{C}+\frac{\arccos(C \sqrt{v_0})}{C^2}.
\end{align*}
To this we must add
\begin{align*}
x_1 - x_{\mbox{\tiny apex}}&= \frac{\sqrt{v_1} \sqrt{1-C^2 v_1}}{C} -\frac{\arcsin(C \sqrt{v_1})}{C^2}+ \frac{\pi}{2 C^2} \\
&=\frac{\sqrt{v_1} \sqrt{1-C^2 v_1}}{C}+\frac{\arccos(C \sqrt{v_1})}{C^2},
\end{align*}
so, in total we obtain condition (\ref{E:nes4}) for $C$.
In the same way we must add the corresponding distance formulas to get formula (\ref{E:nes3}).

This completes the proof of Theorem \ref{T:Hd1}.

We will say that the points $(x_0,y_0)$ and
$(x_1,y_1)$ are $C$-close provided that the inequality in (\ref{E:c-reg1}) holds.
Similarly, if the inequality in (\ref{E:c-reg2}) holds, then we will say that the points are
$C$-far. The equations in (\ref{E:nes2}) and (\ref{E:nes4}) will be called the $C$-equations,
while the formulas in (\ref{E:nes1}) and (\ref{E:nes3}) will be called the $C$-formulas.
\begin{remark}\label{R:susp} \rm
Two different formulas appear in Theorem \ref{T:Hd1} because certain subtleties which underlie the geometry of the cycloid have to be dealt with. Note that formula (\ref{E:nes1}) was suggested as the Heston distance formula in the book \cite{H-L} by P. Henry-Labord\'{e}re
(see formula (6.66) in \cite{H-L}). However, formula (\ref{E:nes1}) holds only in the close-point regime
(see part (i) of Theorem \ref{T:Hd1}) and has to be replaced by formula (\ref{E:nes2}) in the far-point regime. The presence of two different regimes was not taken into account in \cite{H-L}.
\end{remark}

It follows from part (ii) of Theorem \ref{T:Hd1} that
\begin{equation}
d_H((x_0,0),(x_1,0))=2\sqrt{\pi|x_1-x_0|}.
\label{E:osh}
\end{equation}
The previous formula describes the Heston distance between any two points on the
boundary of the Heston half-plane.
\section{The toy Heston model and the Grushin model}\label{S:GtH}
Consider the following stochastic model:
\begin{equation}
\left\{\begin{array}{ll}
dG_t=H_tdW_t \\
dH_t=\frac{1}{2}dZ_t.
\end{array}
\right.
\label{E:Gru-shin}
\end{equation}
We will call the model described by (\ref{E:Gru-shin}) the Grushin model because the Laplace operator associated with it
is a special Grushin operator given by
\begin{equation}
L=\frac{1}{4}\frac{\partial^2}{\partial y^2}+y^2\frac{\partial^2}{\partial x^2}.
\label{E:Gruo}
\end{equation}
The heat kernel for
the operator $L$ will be denoted by $p_t^G((x_0,y_0),(x_1,y_1))$ where $t> 0$,
$(x_0,y_0)\in\mathbb{R}^2$, and $(x_1,y_1)\in\mathbb{R}^2$, and the sub-Riemannian distance
on $\mathbb{R}^2$ (a Carnot-Carath\'{e}odory disrance), corresponding to the Grushin model will
be denoted by $d_G$. The plane $\mathbb{R}^2$ equipped with the distance $d_G$ is called the Grushin plane. The heat kernel $p^G$ for the Grushin plane satisfies the following partial differential equation :
\begin{equation}
\frac{\partial p^G_t}{\partial t}=\frac{1}{8}\frac{\partial^2p^G_t}{\partial y^2}+\frac{1}{2}y^2\frac{\partial^2p^G_t}{\partial x^2},
\label{E:heatk}
\end{equation}
with the initial condition given by
$$
p^G_0\left((x_0,y_0),(x_1,y_1)\right)=\delta_{x_0}(x_1)\delta_{y_0}(y_1)
$$
More information on the geometry of the Grushin plane can be found in
\cite{CCFI,CL,CCGGL,D,P1,P2}. Stochastic methods which are used in the study of Grushin type structures are discussed in \cite{CCHL}.

Suppose the process $(G,H)$ is the solution to the system in (\ref{E:Gru-shin}) with the initial conditions
$g_0$ and $h_0$, respectively. Then the process $(X,Y)$, where $X=G$ and $Y=H^2$, solves the following
system of stochastic differential equations:
\begin{equation}
\left\{\begin{array}{ll}
dX_t=\sqrt{Y_t}d\widetilde{W}_t \\
dY_t=\frac{1}{4}dt+\sqrt{Y_t}d\widetilde{Z}_t
\end{array}
\right.
\label{E:Gru-shkin}
\end{equation}
with initial conditions $g_0$ and $h_0^2$. In (\ref{E:Gru-shkin}), the processes $\widetilde{W}$ and $\widetilde{Z}$ are new standard Brownian motions defined by
$\widetilde{W}_t=\mbox{\rm sign}(H_t)dW_t$ and $\widetilde{Z}_t=\mbox{\rm sign}(H_t)dZ_t$. We will call
the stochastic model described by (\ref{E:Gru-shkin}) the toy Heston model. In this section, the Heston distance will be analyzed by using the following formula relating the Heston and the Grushin distances:
\begin{equation}
d_H((x_0,v_0),(x_1,v_1))=d_G((x_0,\sqrt{v_0}),(x_1,\sqrt{v_1})),
\label{E:GruHes}
\end{equation}
for all points $A=(x_0,v_0)\in{\cal H}$ and $B=(x_1,v_1)\in{\cal H}$.
It is not hard to prove equality (\ref{E:GruHes}) when $A$ and $B$ belong to ${\cal H}^{\circ}$, and then extend the equality to ${\cal H}$ by continuity. The proof for ${\cal H}^{\circ}$ is based on the fact that the length minimizing Heston and Grushin geodesics for $A\in{\cal H}^{\circ}$ and $B\in{\cal H}^{\circ}$ are entirely contained in ${\cal H}^{\circ}$. We leave
filling in the details as an exercise for the reader. It follows from (\ref{E:GruHes}) and
(\ref{E:twos1}) that
\begin{equation}
D(x_0,v_0,x_1,v_1)\le d_G((x_0,\sqrt{v_0}),(x_1,\sqrt{v_1}))\le 12D(x_0,v_0,x_1,v_1)
\label{E:twos2}
\end{equation}
for all $(x_0,v_0)\in{\cal H}$ and $(x_1,v_1)\in{\cal H}$,
where the function $D$ is given by (\ref{E:twos3}).

We will next formulate a statement which provides an alternative formula for the Heston distance.
\begin{theorem}\label{T:Hd2}
For any two points $(x_0,v_0)\in{\cal H}$ and $(x_1,v_1)\in{\cal H}$ such that at least one of
them is not on the boundary, the following formula holds:
\begin{equation}
d_H\left((x_0,v_0),(x_1,v_1)\right)=
\frac{\hat{\delta}}{\sin\left(\frac{\hat{\delta}}{2}\right)}
\sqrt{v_1+v_0-2\sqrt{v_1v_0}\cos\left(\frac{\hat{\delta}}{2}\right)},
\label{E:nes5}
\end{equation}
where $\hat{\delta}=\hat{\delta}((x_0,v_0),(x_1,v_1))$
is the unique solution to the equation
\begin{align}
&\frac{\left(v_1+v_0\right)
\left(\delta-\sin(\delta)\right)
-2\sqrt{v_1v_0}\left(\delta\cos\left(\frac{\delta}{2}\right)
-2\sin\left(\frac{\delta}{2}\right)\right)}
{2\sin^2\left(\frac{\delta}{2}\right)}
\nonumber \\
&=x_1-x_0,
\label{E:nes6}
\end{align}
satisfying the condition $-2\pi<\hat{\delta}< 2\pi$.
\end{theorem}
\begin{remark}\label{R:Pau-lat} \rm
In \cite{P1,P2}, M. Paulat established a formula for the sub-Riemannian distance in a slightly different Grushin model given by:
$$
\left\{\begin{array}{ll}
dG_t=H_tdW_t \\
dH_t=dZ_t.
\end{array}
\right.
$$
Paulat's formula is equivalent to (\ref{E:nes5})
(one formula can be obtained from the other using (\ref{E:GruHes})). The ideas used in the proof of formula
(\ref{E:nes5}) are completely different from those employed in \cite{P1,P2}. Paulat analyzes sub-Riemmanian geodesics in his proof, while the techniques used in the present paper are more analytical. In addition,
the proof of Theorem \ref{T:Hd2} contains several new results, e.g., a partial large devation principle for the Grushin model. We would like to thank M. Paulat for sending us his dissertation \cite{P2}.
\end{remark}

Note that there are two distance formulas in Theorem \ref{T:Hd1} (in the close point regime and
in the far point regime), while Theorem \ref{T:Hd2} contains only one distance formula.
An interesting fact is that
in the $\delta$-environment, there is a special two-set partition of
${\cal H}\times{\cal H}$ hidden in the background.
\begin{definition}\label{D:rational}
We will say that the points $(x_0,v_0)$ and
$(x_1,v_1)$ are $\delta$-close provided that
\begin{equation}
|x_1 - x_0| \leq \frac{\pi}{2}\left(v_1+v_0\right)+2\sqrt{v_1v_0},
\label{E:qe1}
\end{equation}
and $\delta$-far if
\begin{equation}
 | x_1 - x_0| >\frac{\pi}{2}\left(v_1+v_0\right)+2\sqrt{v_1v_0}.
\label{E:qe2}
\end{equation}
\end{definition}
\begin{remark}\label{R:farr} \rm
In terms of the parameter $\hat{\delta}$, the description of the close-point $\delta$-regime and the far-point $\delta$-regime
is $0\le|\hat{\delta}|\le\pi$ and $\pi<|\hat{\delta}|<2\pi$, respectively.
\end{remark}
\begin{remark}\label{R:prop} \rm
It follows from Theorem \ref{T:Hd1} or Theorem \ref{T:Hd2} that
$$
d_H\left((x_0,v_0),(x_1,v_1)\right)=d_H\left((x_1,v_0),(x_0,v_1)\right)
$$
and
$$
d_H\left((x_0,v_0),(x_1,v_1)\right)=d_H\left((x_0,v_1),(x_1,v_0)\right).
$$
\end{remark}

The reasons, why the two regimes in Definition \ref{D:rational} are introduced, are rather subtle.
It will be shown in the next sections that the close-point $\delta$-regime
describes those pairs of points, for which formula (\ref{E:nes5}) can be obtained by analytical methods.
The far point $\delta$-regime is a proper part of the far point $C$-regime,
and formula (\ref{E:nes5}) in the far point regime can be established using formula (\ref{E:nes3}) (see Lemma \ref{L:hg} below).

We will next derive formula (\ref{E:osh}) from Theorem \ref{T:Hd2}. With no loss of generality we can assume $x_0< x_1$. Let $\varepsilon> 0$, and take $v_0=0$, $v_1=\varepsilon$. Then Theorem \ref{T:Hd2} implies
that
$$
d_H((x_0,0),(x_1,\varepsilon))
=\hat{\delta}\left[\sin\left(\frac{\hat{\delta}}{2}\right)\right]^{-1}\sqrt{\varepsilon}
$$
where
$$
\frac{\varepsilon(\hat{\delta}-\sin\hat{\delta})}{2\sin^2\frac{\hat{\delta}}{2}}=x_1-x_0.
$$
Therefore,
\begin{equation}
d_H((x_0,0),(x_1,\varepsilon))
=\frac{\sqrt{2(x_1-x_0)}\hat{\delta}}{\sqrt{\hat{\delta}-\sin\hat{\delta}}}.
\label{E:oshi}
\end{equation}
It is not hard to see that $\hat{\delta}\rightarrow 2\pi$ as $\varepsilon\downarrow 0$.
Taking the limit as
$\varepsilon\downarrow 0$ in formula (\ref{E:oshi}), we obtain formula (\ref{E:osh}).
\section{Solvability and Convexity}\label{S:cas}
In this section we discuss the unique solvability of the $C$-equations and of the $\delta$-equation.
Let us start with the $C$-close point regime. It is clear from the definition of the function $F$ in (\ref{E:C-funct}) that to study the unique solvability
of equation (\ref{E:nes2}), it suffices to assume $x_1> x_0$ and $v_1> v_0$. Then the equation becomes
\begin{equation}
F(v_0,v_1,C)=x_1-x_0,
\label{E:beeq}
\end{equation}
and we have to solve it on the interval $0< C<v_1^{-\frac{1}{2}}$. Note that in the $C$-close point regime
we have $x_1-x_0\le F(v_0,v_1,v_1^{-\frac{1}{2}})$, and the function $F$ maps the interval $[0,v_1^{-\frac{1}{2}})$ onto the interval $[0,F(v_0,v_1,v_1^{-\frac{1}{2}}))$ (the monotonicity of the function $F$ follows fom the next lemma).
\begin{lemma}\label{L:hidden}
For fixed $v_1> v_0$, the function $C\mapsto F(v_0,v_1,C)$ is strictly increasing
on the interval $[0,v_1^{-\frac{1}{2}})$ and
convex on the interval $(0,v_1^{-\frac{1}{2}})$.
\end{lemma}

\it Proof. \rm
Consider the function
\beaa
 F_1(v, C)=  \frac{\arcsin(C \sqrt{v})}{C^2}- \frac{\sqrt{v} \sqrt{1-C^2 v}}{C}.
 \eeaa
By definition, the function $F$ satisfies
\begin{equation}
F(v_0, v_1, C)= F_1(v_1, C) - F_1(v_0, C).
\label{E:byd}
\end{equation}
Therefore
 \bea
&&  \frac{\partial^2 F(v_0, v_1, C)}{\partial C^2} = \frac{\partial^2 F_1}{\partial C^2}(v_1, C) - \frac{\partial^2 F_1}{\partial C^2}(v_0, C) \nn
 \\
 && = \int_{v_0}^{v_1}  \frac{\partial^3 F_1}{\partial v \partial C^2 }(v, C) dy.
 \label{identity}
 \eea

By differentiating the function $F_1$, we get
\begin{align}
\frac{\partial F_1}{\partial C}&=-\frac{2\arcsin(C\sqrt{v})}{C^3}+\frac{\sqrt{v}}{C^2\sqrt{1-C^2v}}
\nonumber \\
&\quad+\frac{\sqrt{v}\sqrt{1-C^2v}}{C^2}+\frac{v^{\frac{3}{2}}}{\sqrt{1-C^2v}}
\label{E:didi1}
\end{align}
and
\bea
 \frac{\partial^2 F_1}{\partial C^2}(v, C)=\frac{2\sqrt{v}\left(-3+4C^2v\right)}
 {C^3\left(1-C^2v\right)^{\frac{3}{2}}}+\frac{6\arcsin(C\sqrt{v})}{C^4}.
 \label{remarkable2}
 \eea
We also have
 \bea
  \frac{\partial^3 F_1}{\partial C^2 \partial v}(v, C)
   = \frac{3Cv^{\frac{3}{2}}}{\left(1-C^2 v \right)^{5/2}}.
  \label{remarkable1}
 \eea
This is quite remarkable because
if we stop at the second derivative we have the more
complicated expression given in (\ref{remarkable2}).
Now, it is not hard to see, using (\ref{identity}) and (\ref{remarkable1}), that
\begin{equation}
\frac{\partial^2 F(v_0, v_1, C)}{\partial C^2}> 0
\label{E:didi3}
\end{equation}
for all $C\in\left(0,v_1^{-\frac{1}{2}}\right)$,
and hence
the convexity statement in Lemma \ref{L:hidden} holds.

We will next prove that the function $F$ is increasing. Using (\ref{E:byd}) and (\ref{E:didi1}),
and making tedious but straightforward computations, we obtain
\begin{equation}
 \frac{\partial F}{\partial C}=: \lim\limits_{C\to 0} \frac{\partial F}{\partial C}(v_0, v_1, C)
= \frac23( v_1^\frac32 - v_0^\frac32) > 0.
\label{E:didi2}
\end{equation}
Now the fact that the function $F$ is increasing follows from
(\ref{E:didi3}), (\ref{E:didi2}), and the equality
$$
\frac{\partial F}{\partial C}(v_0,v_1,C)=\int_0^C\frac{\partial^2 F}{\partial C^2}(v_0, v_1, u)du
+\frac{\partial F}{\partial C}(v_0,v_1,0).
$$

This completes the proof of Lemma \ref{L:hidden}.

It follows from Lemma \ref{L:hidden} that equation (\ref{E:beeq}) is uniquely solvable for all pairs
of points in the Heston half-plane which are $C$-close.

Our next goal is to prove a similar result for any pair of points $(x_0,v_0)$ and $(x_1,v_1)$
in the Heston half-plane which are $C$-far. With no loss of generality we assume
$x_0< x_1$ and $v_0< v_1$.
Recall that the $C$-far point regime is described by the following inequality:
$$
x_1 - x_0\ge  F\left(v_0,v_1,\frac{1}{\sqrt{v_1}}\right).
$$
The $C$-equation in the far point regime is as follows:
\begin{equation}
\widetilde{F}(v_0,v_1,C)=x_1-x_0,
\label{E:equador}
\end{equation}
where
\begin{align}
\widetilde{F}(v_0,v_1,C)&=\frac{\sqrt{ v_0}\sqrt{1-C^2 v_0}}{C}+\frac{\arccos(C \sqrt{v_0})}{C^2}
\nonumber \\
&\quad+ \frac{\sqrt{v_1} \sqrt{1-C^2 v_1}}{C}+\frac{\arccos(C \sqrt{v_1})}{C^2},
\label{E:equador1}
\end{align}
and we are looking for the solution $C^{*}$ satisfying $0< C^{*}< v_1^{-\frac{1}{2}}$.

The function $\widetilde{F}$ is decreasing on the interval $\left(0,v_1^{-\frac{1}{2}}\right)$
(see Lemma \ref{L:nohom} below), and maps it onto the infinite interval
$\left(F\left(v_0,v_1,\frac{1}{\sqrt{v_1}}\right),\infty\right)$. Indeed, it follows from the definition
of $\widetilde{F}$ that $\displaystyle{\lim_{C\downarrow 0}\widetilde{F}(v_0,v_1,C)=\infty}$.
We also have
\begin{align*}
&\lim_{C\uparrow v_1^{-\frac{1}{2}}}\widetilde{F}(v_0,v_1,C)=\sqrt{v_0v_1}\sqrt{1-\frac{v_0}{v_1}}
+v_1\arccos\sqrt{\frac{v_0}{v_1}} \\
&=\sqrt{v_0v_1}\sqrt{1-\frac{v_0}{v_1}}+\frac{\pi}{2}v_1-v_1\arcsin\sqrt{\frac{v_0}{v_1}}
=F\left(v_0,v_1,\frac{1}{\sqrt{v_1}}\right).
\end{align*}
It is clear from the previous discussion that equation (\ref{E:equador}) is unquely solvable
in the $C$-far point regime and the solution $C^{*}$ satisfies $0< C^{*}< v_1^{-\frac{1}{2}}$.
\begin{lemma}\label{L:nohom}
Let $v_0< v_1$. Then the function $\widetilde{F}$ defined by (\ref{E:equador1})
is decreasing on the interval $\left(0,v_1^{-\frac{1}{2}}\right)$. It is locally convex near the
point $C=0$ and locally concave near the point $C=v_1^{-\frac{1}{2}}$.
\end{lemma}

\it Proof. \rm Put
\begin{equation}
\widetilde{F}_1(v,C)=\frac{\sqrt{v}\sqrt{1-C^2v}}{C}+\frac{\arccos(C\sqrt{v})}{C^2}.
\label{E:diffeo0}
\end{equation}
Then we have
\begin{equation}
\widetilde{F}(v_0,v_1,C)=\widetilde{F}_1(v_0,C)+\widetilde{F}(v_1,C).
\label{E:diffeo00}
\end{equation}
By differentiating the function $\widetilde{F}$ twice and simplifying, we obtain
\begin{equation}
\frac{\partial\widetilde{F}_1}{\partial C}(v,C)=-\left[\frac{2\sqrt{v}}{C^2\sqrt{1-C^2v}}
+\frac{2\arccos(C\sqrt{v})}{C^3}\right]
\label{E:difeo1}
\end{equation}
and
\begin{equation}
\frac{\partial^2\widetilde{F}_1}{\partial C^2}(v,C)=\frac{6\sqrt{v}-8v^{\frac{3}{2}}C^2}{C^3(1-C^2v)
^{\frac{3}{2}}}+\frac{6\arccos(C\sqrt{v})}{C^4}.
\label{E:diffeo2}
\end{equation}

It is clear from (\ref{E:difeo1}) that the function $C\mapsto\widetilde{F}_1(v,C)$ decreases. Moreover,
formula (\ref{E:diffeo2}) shows that
$$
\lim_{u\downarrow 0}\frac{\partial^2\widetilde{F}_1}{\partial C^2}(v,u)=\infty
$$
and
$$
\lim_{u\uparrow v^{-\frac{1}{2}}}\frac{\partial^2\widetilde{F}_1}{\partial C^2}(v,u)=-\infty.
$$
Analyzing the previous equalities and taking into account (\ref{E:diffeo00}), we see that Lemma
\ref{L:nohom} holds.

Let us next consider the equation in (\ref{E:nes6}). We fix $x_0\in\mathbb{R}$, $x_1\in\mathbb{R}$,
$v_0\ge 0$, $v_1\ge 0$ and assume that at least one of the numbers $v_0$ and $v_1$ is different from zero. Denote the function on the left-hand side of (\ref{E:nes6}) by $f(\delta)$.
Then the equation in (\ref{E:nes6}) can be rewritten as follows:
\begin{equation}
f(\delta)=x_1-x_0.
\label{E:can10}
\end{equation}
We have
\begin{equation}
f(\delta)=f(\delta,v_0,v_1)=\frac{1}{2}A(\delta)B(\delta)
\label{E:cand0}
\end{equation}
where
\begin{equation}
A(\delta)=A(\delta,v_0,v_1)=\sin^{-2}\left(\frac{\delta}{2}\right)
\label{E:can20}
\end{equation}
and
\begin{align}
&B(\delta)=B(\delta,v_0,v_1)=\left(v_1+v_0\right)
\left(\delta-\sin(\delta)\right) \nonumber \\
&\quad-2\sqrt{v_1v_0}\left(\delta\cos\left(\frac{\delta}{2}\right)
-2\sin\left(\frac{\delta}{2}\right)\right).
\label{E:can30}
\end{align}
The value of the function $f$ at $\delta=0$ is given by
\begin{equation}
f(0)=\frac{1}{2}\lim_{\delta\rightarrow 0}A(\delta)B(\delta)=0.
\label{E:atzz}
\end{equation}

The function $f$ is continuous and odd on the interval $(-2\pi,2\pi)$. In order to prove that the equation in (\ref{E:nes6}) is uniquely solvable on $(-2\pi,2\pi)$, it suffices to assume that $x_1> x_0$ and look for the unique solution belonging to the interval $(0,2\pi)$.
We will next show that the function $f$ is positive
on the interval $(0,2\pi)$. Indeed, the functions $\delta\mapsto\delta-\sin(\delta)$ and
$\delta\mapsto\delta\cos\left(\frac{\delta}{2}\right)-2\sin\left(\frac{\delta}{2}\right)$ are equal to zero at $\delta=0$. Moreover, the former function is positive on $(0,2\pi)$, while the latter one is decreasing (differentiate!) and hence negative on
$(0,2\pi)$. It follows from (\ref{E:can30}) that $B(\delta)> 0$ for all $\delta\in(0,2\pi)$. This shows that the function
$f$ is positive on the interval $(0,2\pi)$.

We also have
$f(\pi)=\frac{1}{2}\left[\pi\left(v_1+v_0\right)+4\sqrt{v_1v_0}\right]$.
\begin{lemma}\label{L:cancan0}
The function $f$ is strictly increasing and convex on the interval $(0,2\pi)$. Moreover, it maps
$(0,2\pi)$ onto $(0,\infty)$.
\end{lemma}
\begin{remark} \rm Lemma \ref{L:cancan0} implies that there exists the unique solution to the equation in (\ref{E:nes6}) belonging to the interval $(-2\pi,2\pi)$.
\end{remark}
\it Proof. \rm
For all $\delta\in(0,2\pi)$, we have
\begin{equation}
A^{\prime}(\delta)=-\sin^{-3}\left(\frac{\delta}{2}\right)\cos\left(\frac{\delta}{2}\right)
\label{E:can60}
\end{equation}
and
\begin{equation}
A^{\prime\prime}(\delta)=\frac{2+\cos(\delta)}{2\sin^4\left(\frac{\delta}{2}\right)},
\label{E:can70}
\end{equation}
\begin{equation}
B^{\prime}(\delta)=\left(v_1+v_0\right)(1-\cos(\delta))+\sqrt{v_1v_0}
\delta\sin\left(\frac{\delta}{2}\right),
\label{E:can80}
\end{equation}
and
\begin{equation}
B^{\prime\prime}(\delta)=\left(v_1+v_0\right)\sin(\delta)+\sqrt{v_1v_0}
\left(\sin\left(\frac{\delta}{2}\right)
+\frac{1}{2}\delta\cos\left(\frac{\delta}{2}\right)\right).
\label{E:can90}
\end{equation}
Next, using the product rule, l'H\^{o}pital's rule, and the formulas above, we see that
\begin{equation}
\lim_{\delta\downarrow 0}f^{\prime}(\delta)=\frac{1}{3}\left(v_1+v_0\right)+\frac{1}{3}\sqrt{v_1v_0}.
\label{E:can100}
\end{equation}
Now, (\ref{E:atzz}), (\ref{E:can100}), and the fact that the function $f$ is odd on $(-2\pi,2\pi)$ imply that $f$ is differentiable at $\delta=0$.

Our goal is to show that
\begin{equation}
f^{\prime\prime}(\delta)> 0\quad\mbox{for}\quad\delta\in(0,2\pi).
\label{E:sos10}
\end{equation}
It is not hard to see that if (\ref{E:sos10}) holds, then the function $f$ is increasing on $(0,2\pi)$. Indeed (\ref{E:sos10}) implies that for all
$0<\varepsilon<\delta<2\pi$,
$$
f^{\prime}(\delta)=f^{\prime}(\varepsilon)+\int_{\varepsilon}^{\delta}f^{\prime\prime}(u)du
>f^{\prime}(\varepsilon).
$$
Therefore, (\ref{E:can100}) shows that the derivative of the function $f$ is positive on $(0,2\pi)$. Now the continuity of $f$ on $[0,2\pi)$
implies that the function $f$ is increasing on $[0,2\pi)$.

Our next goal is to prove (\ref{E:sos10}). We have
\begin{equation}
(AB)^{\prime\prime}=A^{\prime\prime}B+2A^{\prime}B^{\prime}+AB^{\prime\prime}.
\label{E:sos20}
\end{equation}
Using (\ref{E:can20}), (\ref{E:can30}), (\ref{E:can60}), (\ref{E:can70}), (\ref{E:can80}), and (\ref{E:can90}), we obtain
\begin{align*}
&A^{\prime\prime}B=\frac{(2+\cos(\delta))}{2\sin^4\left(\frac{\delta}{2}\right)} \\
&\quad\left[\left(v_1+v_0\right)(\delta-\sin(\delta))
-2\sqrt{v_1v_0}\left(\delta\cos\left(\frac{\delta}{2}\right)
-2\sin\left(\frac{\delta}{2}\right)\right)\right],
\end{align*}
$$
2A^{\prime}B^{\prime}=-\frac{2\cos\left(\frac{\delta}{2}\right)\left[\left(v_1+v_0\right)(1-\cos(\delta))
+\sqrt{v_1v_0}\delta\sin\left(\frac{\delta}{2}\right)\right]}{\sin^3\left(\frac{\delta}{2}\right)},
$$
and
$$
AB^{\prime\prime}=\frac{\left(v_1+v_0\right)\sin(\delta)+\sqrt{v_1v_0}
\left(\sin\left(\frac{\delta}{2}\right)+
\frac{1}{2}\delta\cos\left(\frac{\delta}{2}\right)\right)}{\sin^2\left(\frac{\delta}{2}\right)}.
$$

It is not hard to see using (\ref{E:sos20}) that (\ref{E:sos10}) holds if the function
\begin{align*}
&J(\delta)=(2+\cos(\delta)) \\
&\times\left[\left(v_1+v_0\right)(\delta-\sin(\delta))
-2\sqrt{v_1v_0}\left(\delta\cos\left(\frac{\delta}{2}\right)-2\sin\left(\frac{\delta}{2}\right)\right)\right]  \\
&-4\sin\left(\frac{\delta}{2}\right)\cos\left(\frac{\delta}{2}\right)\left[\left(v_1+v_0\right)(1-\cos(\delta))
+\sqrt{v_1v_0}\delta\sin\left(\frac{\delta}{2}\right)\right] \\
&+2\sin^2\left(\frac{\delta}{2}\right)\left[\left(v_1+v_0\right)\sin(\delta)+\sqrt{v_1v_0}\left(\sin\left(\frac{\delta}{2}\right)+
\frac{1}{2}\delta\cos\left(\frac{\delta}{2}\right)\right)\right]
\end{align*}
is positive. The previous function can be represented as follows:
$$
J(\delta)=\left(v_1+v_0\right)C_1(\delta)+\sqrt{v_1v_0}C_2(\delta),
$$
where
$$
C_1(\delta)=2\delta+\delta\cos(\delta)-3\sin(\delta)
$$
and
\begin{align}
C_2(\delta)&=6\sin\left(\frac{\delta}{2}\right)\left(1+\cos^2\left(\frac{\delta}{2}\right)\right)
\nonumber \\
&\quad-\delta\cos\left(\frac{\delta}{2}\right)
\left(5+\cos^2\left(\frac{\delta}{2}\right)\right).
\label{E:inin0}
\end{align}
Now it is clear that it suffices to show that $C_1(\delta)> 0$ and $C_2(\delta)> 0$ for all $\delta\in(0,2\pi)$.

We have
$$
C_1(0)=C_1^{\prime}(0)=C_1^{\prime\prime}(0)=0
$$
and
$$
C^{\prime\prime\prime}(\delta)=\delta\sin(\delta)> 0
$$
for all $\delta\in(0,2\pi)$. Therefore $C_1(\delta)> 0$ on the interval $(0,2\pi)$.

In order to prove that $C_2(\delta)> 0$ for all $0<\delta<2\pi$, we reason as follows.
First note that (\ref{E:inin0}) implies the inequality $C_2(\delta)> 0$ for all
$\delta\in[\pi,2\pi)$. Therefore, we can restrict ourselves to the case where $0<\delta<\pi$.
The following estimates, which can be easily derived using Taylor expansions of the
sine and cosine functions, will be needed in the proof below:
\begin{equation}
\sin\left(\frac{\delta}{2}\right)>\frac{\delta}{2}-\frac{\delta^3}{48},
\label{E:T10}
\end{equation}
\begin{equation}
\cos\left(\frac{\delta}{2}\right)< 1-\frac{\delta^2}{8}+\frac{\delta^4}{384},
\label{E:T20}
\end{equation}
and
\begin{equation}
\sin\delta<\delta-\frac{\delta^3}{6}+\frac{\delta^5}{120}
\label{E:T30}
\end{equation}
for all $\delta\in(0,\pi)$.

Using (\ref{E:T10}) and (\ref{E:T20}), we obtain
\begin{align}
&C_2(\delta)=6\sin\left(\frac{\delta}{2}\right)\left(1+\cos^2\left(\frac{\delta}{2}\right)\right)-\delta\cos\left(\frac{\delta}{2}\right)
\left(5+\cos^2\left(\frac{\delta}{2}\right)\right) \nonumber \\
&>\delta\left[\left(3-\frac{\delta^2}{8}\right)\left(1+\cos^2\left(\frac{\delta}{2}\right)\right)-\left(
1-\frac{\delta^2}{8}+\frac{\delta^4}{384}\right)\left(5+\cos^2\left(\frac{\delta}{2}\right)\right)\right] \nonumber \\
&=\delta\left[\frac{\delta^2}{2}+2\cos^2\left(\frac{\delta}{2}\right)-2-\frac{\delta^4}{384}
\left(5+\cos^2\left(\frac{\delta}{2}\right)\right)\right] \nonumber \\
&\ge\delta\left[\frac{\delta^2}{2}-\frac{\delta^4}{64}-2\sin^2\left(\frac{\delta}{2}\right)\right].
\label{E:sis20}
\end{align}

It will be shown next that the function
$$
h(\delta)=\frac{\delta^2}{2}-\frac{\delta^4}{64}-2\sin^2\left(\frac{\delta}{2}\right)
$$
is positive on the interval $(0,\pi)$. Indeed, $h(0)=0$ and
$$
h^{\prime}(\delta)=\delta-\frac{\delta^3}{16}-\sin\delta.
$$
Now using (\ref{E:T30}) we get
$$
h^{\prime}(\delta)>\delta^3\left(\frac{5}{48}-\frac{\delta^2}{120}\right)> 0
$$
since $\delta\in(0,\pi)$.
It follows that $C_2(\delta)> 0$ for $0<\delta<\pi$. Finally, it is easy to check, using the definition of the function $f$, that $f$ maps
$(0,2\pi)$ onto $(0,\infty)$.

This completes the proof of Lemma \ref{L:cancan0}.

\begin{lemma}\label{L:regimes}
Suppose the points $(x_0,v_0)\in{\cal H}$ and $(x_1,v_1)\in{\cal H}$ are $\delta$-far. Then they are $C$-far.
\end{lemma}

\it Proof. \rm It suffices to prove the lemma in the case where $v_1> 0$, $v_0\ge 0$, and $v_1\ge v_0$.
The assumption in the formulation of Lemma \ref{L:regimes} means that inequality
(\ref{E:qe2}) holds. To prove that the points are $C$-far, we have to establish estimate
(\ref{E:c-reg2}). For $v_1\ge v_0$, this estimate is as follows:
\begin{equation}
|x_1-x_0|>\frac{\pi}{2}v_1+\sqrt{v_1v_0-v_0^2}-v_1\arcsin\sqrt{\frac{v_0}{v_1}}.
\label{E:far1}
\end{equation}
It is easy to see that condition (\ref{E:qe2}) implies condition (\ref{E:far1}).

This concludes the proof of Lemma \ref{L:regimes}.

Let us define the following function:
\begin{equation}
 f(u,v_0,v_1)=\frac{\sin\frac{u}{2}}{\sqrt{v_1+v_0-2\sqrt{v_1v_0}\cos\frac{u}{2}}}.
 \label{E:funnuf}
\end{equation}
It is assumed in (\ref{E:funnuf}) that $u\in(\pi,2\pi)$, $v_0\ge 0$, and $v_1\ge 0$.
We also exclude the case where $v_0=v_1=0$. It is not hard to see that for fixed $v_0$ and $v_1$, the function
$u\mapsto f(u,v_0,v_1)$ is strictly decreasing, continuous, and
\begin{equation}
f:(\pi,2\pi)\mapsto\left(0,\frac{1}{\sqrt{v_1+v_0}}\right)
\label{E:ontto}
\end{equation}
(the mapping is onto).

Suppose the points $(x_0,v_0)\in{\cal H}$ and $(x_1,v_1)\in{\cal H}$ are $\delta$-far.
By Lemma \ref{L:regimes}, these points are also $C$-far. It is natural to ask
wheter there exists a relation between the numbers $\delta^{*}$ and $C^{*}$, corresponding to the given
points. The next statement answers the previous question.
\begin{lemma}\label{L:connection}
If $(x_0,v_0)\in{\cal H}$ and $(x_1,v_1)\in{\cal H}$ are $\delta$-far, then
\begin{equation}
C^{*}((x_0,v_0),(x_1,v_1))=f\left(|\hat{\delta}((x_0,v_0),(x_1,v_1))|,v_0,v_1\right),
\label{E:connectt1}
\end{equation}
where the function $f$ is defined by (\ref{E:funnuf}).
 \end{lemma}

\it Proof. \rm Since the points are $\delta$-far, we have $\pi<|\hat{\delta}|< 2\pi$.
With no loss of generality, we can assume that $\pi<\hat{\delta}< 2\pi$, $x_0\le x_1$, and $v_0\le v_1$.
Note that  condition (\ref{E:ontto}) implies that all the expressions appearing in the proof of Lemma
\ref{L:connection} are real numbers.

Let us denote $\alpha=\sqrt{\frac{v_0}{v_1}}$, and recall that the number $\hat{\delta}$ is the unique
solution to equation (\ref{E:nes6}). Then
\begin{align}
&x_1-x_0 \nonumber \\
&=v_1\frac{\left(1+\alpha^2\right)
\left(\hat{\delta}-\sin(\hat{\delta})\right)
-2\alpha\left(\hat{\delta}\cos\left(\frac{1}{2}\hat{\delta}\right)
-2\sin\left(\frac{1}{2}\hat{\delta}\right)\right)}
{2\sin^2\left(\frac{1}{2}\hat{\delta}\right)}.
\label{E:nestor6}
\end{align}
Put
\begin{align}
\widetilde{C}&=\sqrt{v_1}f\left(\hat{\delta}((x_0,v_0),(x_1,v_1)),v_0,v_1\right) \nonumber \\
&=\frac{\sin\left(\frac{1}{2}\hat{\delta}\right)}
{\sqrt{1+\alpha^2-2\alpha\cos\left(\frac{1}{2}\hat{\delta}\right)}}.
\label{E:wide1}
\end{align}
It is not hard to see that equality (\ref{E:connectt1}) holds if and only if
\begin{align}
&x_1-x_0 \nonumber \\
&=v_1\left[\frac{\alpha\sqrt{1-\alpha^2\widetilde{C}^2}+\sqrt{1-\widetilde{C}^2}}{\widetilde{C}}
+\frac{\arccos\widetilde{C}+\arccos(\alpha\widetilde{C})}{\widetilde{C}^2}\right],
\label{E:wide2}
\end{align}
where $\widetilde{C}$ is defined by (\ref{E:wide1}). Using the addition formula for the inverse cosines,
we see that equality (\ref{E:wide2}) is equivalent to the following equality:
\begin{align}
\frac{x_1-x_0}{v_1}&=\alpha\sqrt{\frac{1}{\widetilde{C}^2}-\alpha^2}
+\sqrt{\frac{1}{\widetilde{C}^2}-1} \nonumber \\
&\quad+\frac{\arccos(\alpha\widetilde{C^2}-\sqrt{1-\widetilde{C}^2}\sqrt{1-\alpha^2\widetilde{C}^2})}
{\widetilde{C}^2}.
\label{E:wide3}
\end{align}

Since $\hat{\delta}$ solves equation (\ref{E:nes6}), we have
\begin{align}
&x_1-x_0 \\
&=v_1\frac{(1+\alpha^2)(\hat{\delta}-\sin\hat{\delta})
+2\alpha\left(2\sin\left(\frac{1}{2}\hat{\delta}\right)
-\hat{\delta}\cos\left(\frac{1}{2}\hat{\delta}\right)\right)}
{2\sin^2\left(\frac{1}{2}\hat{\delta}\right)}.
\label{E:wide4}
\end{align}
We will derive equality (\ref{E:wide3}) from equality (\ref{E:wide4}).
It follows from (\ref{E:wide1}) that
\begin{equation}
\widetilde{C}^2=\frac{1-cos^2\left(\frac{1}{2}\hat{\delta}\right)}{1+\alpha^2-2\alpha\cos\left(\frac{1}{2}\hat{\delta}\right)}.
\label{E:wide5}
\end{equation}
Solving the corresponding quadratic equation for $\cos\left(\frac{1}{2}\hat{\delta}\right)$
and taking into account that $\cos\left(\frac{1}{2}\hat{\delta}\right)< 0$, we get
$$
\cos\left(\frac{1}{2}\hat{\delta}\right)=\alpha\widetilde{C}^2-\sqrt{\alpha^2\widetilde{C}^4
-C^2-\alpha^2\widetilde{C}^2+1}.
$$
Therefore,
\begin{equation}
\cos\left(\frac{1}{2}\hat{\delta}\right)=\alpha\widetilde{C}^2-\sqrt{1-\widetilde{C}^2}\sqrt
{1-\alpha^2\widetilde{C}^2}.
\label{E:connect1}
\end{equation}
Next, using (\ref{E:wide5}) and (\ref{E:connect1}), we see that equality (\ref{E:wide3})
is equivalent to the following:
\begin{align}
&\frac{x_1-x_0}{v_1}=\alpha\sqrt{\frac{1+\alpha^2-2\alpha\cos\left(\frac{1}{2}\hat{\delta}\right)}
{\sin^2\left(\frac{1}{2}\hat{\delta}\right)}-\alpha^2} \nonumber \\
&\quad+\sqrt{\frac{1+\alpha^2-2\alpha\cos\left(\frac{1}{2}\hat{\delta}\right)}
{\sin^2\left(\frac{1}{2}\hat{\delta}\right)}-1}
+\frac{\hat{\delta}(1+\alpha^2-2\alpha\cos\left(\frac{1}{2}\hat{\delta}\right))}
{2\sin^2\left(\frac{1}{2}\hat{\delta}\right)}.
\label{E:wide6}
\end{align}
Making simplifications in (\ref{E:wide6}), we can show that equality
(\ref{E:wide6}) reduces to equality (\ref{E:wide4}).

This completes the proof of Lemma \ref{L:connection}.
\begin{lemma}\label{L:hg}
The $\hat{\delta}$-formula for the Heston distance in the case, where the points
$(x_0,v_0)\in{\cal H}$ and $(x_1,v_1)\in{\cal H}$ are
$\delta$-far, follows from part (ii) of Theorem \ref{T:Hd2}.
\end{lemma}

\it Proof. \rm In the proof of Lemma \ref{L:hg}, we will use the notation
introduced in the proof of Lemma \ref{L:connection}.
Suppose the points $(x_0,v_0)\in{\cal H}$ and $(x_1,v_1)\in{\cal H}$ are
$\delta$-far, and assume that part (ii)
of Theorem \ref{T:Hd2} is valid. By Lemma \ref{L:regimes}, $(x_0,v_0)$
and $(x_1,v_1)$ are $C$-far, and it follows from part
(ii) of Theorem \ref{T:Hd2} that
$$
d_H((x_0, v_0), (x_1, v_1))=2\frac{\arccos(C^* \sqrt{v}_0) +\arccos(C^* \sqrt{v}_1))}{C^*}.
$$
Taking into account (\ref{E:connectt1}), (\ref{E:wide1}), and (\ref{E:connect1}), we see that
\begin{align*}
&d_H((x_0, v_0), (x_1, v_1))=2\sqrt{v_1}\frac{\arccos(\alpha\widetilde{C})
+\arccos(\widetilde{C})}{C^*}
\\
&=2\sqrt{v_1}\frac{\arccos(\alpha\widetilde{C^2}-\sqrt{1-\widetilde{C}^2}
\sqrt{1-\alpha^2\widetilde{C}^2})}{C^*}
=\frac{\sqrt{v_1}\hat{\delta}}{\widetilde{C}}.
\end{align*}
Next, using (\ref{E:wide1}), we obtain
$$
d_H((x_0, v_0), (x_1, v_1))=\frac{\hat{\delta}\sqrt{v_1}\sqrt{1+\alpha^2-2\alpha\cos\left(\frac{1}{2}\hat{\delta}\right)}}
{\sin\left(\frac{1}{2}\hat{\delta}\right)}.
$$
Finally, recalling that $\alpha=\sqrt{\frac{v_0}{v_1}}$, we see that formula (\ref{E:nes5}) holds.

This completes the proof of Lemma \ref{L:hg}.
\section{The limiting cumulant generating function for the Grushin model}\label{S:lcgf}
It is not hard to see using the equation in (\ref{E:heatk}) that the Laplace transform
$\tilde{p}^G_t((x_0,y_0),(w,y_1))$ in the variable $x_1$ of the Grushin transition density $p_T^G$ satisfies the following heat equation with quadratic potential:
$$
\frac{\partial\tilde{p}^G_t}{\partial t}=\frac{1}{8}\frac{\partial^2\tilde{p}^G_t}{\partial y^2}+\frac{1}{2}w^2y^2\tilde{p}^G_t
$$
with the initial condition given by $\tilde{p}^G_0((x_0,y_0),(w,y_1))=\delta_{x_0}(x_1)e^{wy_0}$.
The fundamental solution for such a heat equation is well-known (see, e.g., \cite{CC}, Theorem
10.3). Using this fundamental solution, we get
\begin{align}
&\tilde{p}^G_t(x_0,y_0,w,y_1)=\sqrt{\frac{w}{\pi\sin\left(\frac{wt}{2}\right)}} \nonumber \\
&\quad\exp\left\{-\frac{w\left[-x_0\sin\left(\frac{wt}{2}\right)
+\left(y_1^2+y_0^2\right)\cos\left(\frac{wt}{2}\right)-2y_1y_0\right]}
{\sin\left(\frac{wt}{2}\right)}\right\}.
\label{E:oo}
\end{align}
The function $\tilde{p}^G_t$ given by (\ref{E:oo}) has a removable singularity at $w=0$. The analyticity strip for $\tilde{p}^G_t$ is
given by $-\frac{2\pi}{t}<\Re(w)<\frac{2\pi}{t}$. It is not clear yet what happens outside this strip.

We will next compute the Laplace transform of the function $\tilde{p}^G_t$ in the variable $y_1$, using formula
(\ref{E:oo}). We have
\begin{align*}
&\int_{-\infty}^{\infty}\int_{-\infty}^{\infty}e^{wx_1+\beta y_1}p^G_t((x_0,y_0),(x_1,y_1))dy_1dx_1
\nonumber \\
&=\sqrt{\frac{w}{\pi\sin\left(\frac{wt}{2}\right)}}
\exp\left\{-\frac{w\left(-x_0\sin\left(\frac{wt}{2}\right)+y_0^2\cos\left(\frac{wt}{2}\right)\right)}
{\sin\left(\frac{wt}{2}\right)}\right\} \\
&\quad\int_{-\infty}^{\infty}e^{\beta y_1}
\exp\left\{-\frac{w\left[y_1^2\cos\left(\frac{wt}{2}\right)-2y_1y_0\right]}{\sin\left(\frac{wt}{2}\right)}\right\}dy_1 
\end{align*}
Let us next replace $w$ by $\frac{\delta}{t}$ and $\beta$ by $\frac{\gamma}{t}$ in the previous equality. This gives the following:
\begin{align}
&\int_{-\infty}^{\infty}\int_{-\infty}^{\infty}
\exp\left\{\frac{1}{t}(\delta x_1+\gamma y_1)\right\}p^G_t((x_0,y_0),(x_1,y_1))dy_1dx_1
\nonumber \\
&=\sqrt{\frac{\delta}{\pi t\sin\left(\frac{\delta}{2}\right)}}
\exp\left\{-\frac{\delta\left(-x_0\sin\left(\frac{\delta}{2}\right)+y_0^2\cos\left(\frac{\delta}{2}\right)\right)}
{t\sin\left(\frac{\delta}{2}\right)}\right\} \nonumber \\
&\quad\int_{-\infty}^{\infty}\exp\left\{\frac{\gamma}{t}y_1\right\}
\exp\left\{-\frac{\delta\left[y_1^2\cos\left(\frac{\delta}{2}\right)-2y_1y_0\right]}
{t\sin\left(\frac{\delta}{2}\right)}\right\}dy_1. 
\label{E:bac}
\end{align}
The new restrictions on the parameters are $-2\pi<\delta<2\pi$ and $\gamma\in\mathbb{R}$.

Denote the expression on the left-hand side of (\ref{E:bac}) by $J_t(x_0,\delta,y_0,\gamma)$.
Assume $-\pi<\delta<\pi$ and $\gamma\in\mathbb{R}$. After lengthy but straightforward computations,
we obtain
\begin{align}
&J_t(x_0,\delta,y_0,\gamma)
=\frac{1}{\sqrt{\cos\left(\frac{\delta}{2}\right)}} \nonumber \\
&\quad\times\exp\left\{\frac{1}{t}\frac{\left(4\delta^2y_0^2+\gamma^2\right)\sin\left(\frac{\delta}{2}\right)+4x_0\delta^2\cos\left(\frac{\delta}{2}\right)
+4y_0\gamma\delta}
{4\delta\cos\left(\frac{\delta}{2}\right)}\right\}.
\label{E:errr}
\end{align}
For every pair $(\delta,\gamma)\in\mathbb{R}^2$ and $t> 0$, put
\begin{align}
\Lambda_t(\delta,\gamma)
&=t\log J_t(x_0,\delta,y_0,\gamma)
\label{E:ut1}
\end{align}
and
\begin{equation}
\Lambda(\delta,\gamma)=\lim_{t\rightarrow 0}\Lambda_t(\delta,\gamma)
\label{E:ut2}
\end{equation}
if the limit in (\ref{E:ut2}) exists. Note that the functions $\Lambda_t$ and $\Lambda$ depend on
$x_0$ and $y_0$. The function $\Lambda$ is called the limiting cumulant generating function associated with the Grushin model.

Let us assume that $-\pi<\delta<\pi$ and $\gamma\in\mathbb{R}$. Then,
taking the logarithm of the integral on the left-hand side of (\ref{E:errr}), multiplying by $t$, and sending $t$
to infinity, we obtain
\begin{equation}
\Lambda(\delta,\gamma)=\frac{\left(4\delta^2y_0^2+\gamma^2\right)\sin\left(\frac{\delta}{2}\right)+4x_0\delta^2\cos\left(\frac{\delta}{2}\right)
+4y_0\gamma\delta}
{4\delta\cos\left(\frac{\delta}{2}\right)}.
\label{E:al}
\end{equation}

Let us next assume that
$\delta\in\left(-2\pi,-\pi\right]\bigcup\left[\pi,2\pi\right)$
and $\gamma\in\mathbb{R}$.
Then, using (\ref{E:bac}), we see that
$\Lambda_t(\delta,\gamma)=\infty$ for every $t> 0$. Here we take into account that under the restrictions imposed above,
$$
\delta\cos\left(\frac{\delta}{2}\right)\sin^{-1}\left(\frac{\delta}{2}\right)\le 0.
$$
It follows from (\ref{E:ut2}) that
$\Lambda(\delta,\gamma)=\infty$
for all
$\delta\in\left(-2\pi,-\pi\right]\bigcup\left[\pi,2\pi\right)$
and $\gamma\in\mathbb{R}$. Now, using H\"{o}lder's inequality, we see that
$\Lambda(\delta,\gamma)=\infty$ for all
$\delta\in\left(-\infty,-\pi\right]\bigcup\left[\pi,\infty\right)$.

The limiting cumulant generating function $\Lambda$
is defined everywhere and convex on $\mathbb{R}^2$.
This function is finite on the set
$D=(-\pi,\pi)\times\mathbb{R}$
and identically infinite outside this set.
Moreover, the function $\Lambda$ is continuous on the set
$\left[\mathbb{R}\backslash\left\{-\pi\right\}\backslash\left\{\pi\right\}\right]\times\mathbb{R}$.
It is also continuous on the lines $\delta=\pi$ and $\delta=-\pi$ with the exception of the points $P_1=(\pi,-2\pi y_0)$
and $P_2=(-\pi,-2\pi y_0)$. More precisely, we have
$\Lambda\left(P_1\right)=\Lambda\left(P_2\right)=\infty$. In addition,
\begin{equation}
\liminf_{(\delta,\gamma)\rightarrow P_1,(\delta,\gamma)\in D}\Lambda(\delta,\gamma)=\pi x_0
\label{E:insert1}
\end{equation}
and
\begin{equation}
\liminf_{(\delta,\gamma)\rightarrow P_2,(\delta,\gamma)\in D}\Lambda(\delta,\gamma)=-\pi x_0.
\label{E:insert2}
\end{equation}
Hence, the limiting cumulant generating function $\Lambda$ is lower semi-continuous everywhere in $\mathbb{R}^2$, except at the points $P_1$ and $P_2$.

Using the definition of the function $\Lambda$, we see that for all $(\delta,\gamma)\in D$,
\begin{equation}
\frac{\partial\Lambda}{\partial\delta}
=\frac{4y_0^2\delta^2\left(\delta+\sin\delta\right)
+\gamma^2\left(\delta-\sin\delta\right)
+4\delta^2\gamma y_0\sin\left(\frac{\delta}{2}\right)}{8\delta^2\cos^2\left(\frac{\delta}{2}\right)}+x_0
\label{E:diff1}
\end{equation}
and
\begin{equation}
\frac{\partial\Lambda}{\partial\gamma}=\frac{\gamma\sin\left(\frac{\delta}{2}\right)+2\delta y_0}{2\delta\cos\left(\frac{\delta}{2}\right)}.
\label{E:diff2}
\end{equation}
Hence, the function $\Lambda$ is
continuously differentiable on the set $D$.
However, this function is not steep (the definition of the steepness of a function is given in \cite{DZ}),
Definition 2.3.5). We will next prove the previous statement.

It follows from (\ref{E:diff2}) that
$||\nabla\Lambda\left(\delta,\gamma\right)||\rightarrow\infty$
provided that $(\delta,\gamma)\in D$
and $(\delta,\gamma)\rightarrow (\delta_0,\gamma_0)$ with either $\delta_0=\pi$ and $\gamma_0\neq-2\pi y_0$, or $\delta_0=-\pi$
and $\gamma_0\neq-2\pi y_0$.
The behavior of the gradient at the exceptional points $P_1=\left(\pi,-2\pi y_0\right)$ and $P_2=\left(-\pi,-2\pi y_0\right)$ can be described
using (\ref{E:diff1}), (\ref{E:diff2}), and l'H\^{o}pital's rule. We have
$\displaystyle{\frac{\partial\Lambda}{\partial\delta}\rightarrow
\frac{y_0^2}{2\pi}\left(\pi^2-4\right)+x_0}$
and
$\displaystyle{\frac{\partial\Lambda}{\partial\gamma}\rightarrow-\frac{2y_0}{\pi}}$
as $(\delta,\gamma)\in D$ and $(\delta,\gamma)\rightarrow P_1$.
Therefore,
$$
||\nabla\Lambda\left(\delta,\gamma\right)||\rightarrow\sqrt{\frac{4y_0^2}{\pi^2}+\left(\frac{y_0^2}{\pi^2}\left(\pi^2-4\right)+x_0\right)^2}
$$
as $(\delta,\gamma)\in D$ and $(\delta,\gamma)\rightarrow P_1$. Similarly,
$$
\frac{\partial\Lambda}{\partial\delta}\rightarrow\frac{y_0^2}{2\pi}\left(4-\pi^2\right)+x_0,
\quad\frac{\partial\Lambda}{\partial\gamma}\rightarrow-\frac{2y_0}{\pi},
$$
and
$$
||\nabla\Lambda\left(\delta,\gamma\right)||\rightarrow\sqrt{\frac{4y_0^2}{\pi^2}+\left(\frac{y_0^2}{\pi^2}\left(4-\pi^2\right)+x_0\right)^2}
$$
as $(\delta,\gamma)\in D$ and $(\delta,\gamma)\rightarrow P_2$.
Therefore the steepness condition for the function $\Lambda$ is satisfied everywhere on the boundary of the set $D(\Lambda)$, with the exception of the points
$P_1$ and $P_2$.
\begin{remark}\label{R:nonsteep} \rm
The absence of the lower semi-continuity and of the steepness property for the function $\Lambda$ does
not allow us to use the G\"{a}rtner-Ellis theorem (see Theorem 2.3.6 in \cite{DZ}) to establish the large deviation principle for the Grushin
model. It will be shown below that this principle is valid only in a special regime (see
Theorem \ref{T:sum}).
\end{remark}
\section{Critical points and the Legendre-Fenchel transform}\label{S:how}
Let us consider the Legendre-Fenchel transform $\Lambda^{*}$ of the limiting cumulant generating function $\Lambda$. It is given by
$$
\Lambda^{*}(x_0,y_0,x_1,y_1)=\sup_{\delta,\gamma\in\mathbb{R}}[\langle x_1,\delta\rangle+\langle y_1,\gamma\rangle-\Lambda(\delta,\gamma)].
$$
It is clear that
\begin{equation}
\Lambda^{*}(x_0,y_0,x_1,y_1)=\sup_{-\pi<\delta<\pi,\gamma\in\mathbb{R}}[\langle x_1,\delta\rangle+\langle y_1,\gamma\rangle-\Lambda(\delta,\gamma)].
\label{E:cp1}
\end{equation}

A critical point $(\delta,\gamma)=(\delta^{*},\gamma^{*})$, $-\pi<\delta^{*}<\pi$, for the function
$\langle x_1,\delta\rangle+\langle y_1,\gamma\rangle-\Lambda(\delta,\gamma)$ satisfies
the following system of equations:
\begin{equation}
\frac{\partial\Lambda}{\partial\delta}=x_1,\quad\frac{\partial\Lambda}{\partial\gamma}=y_1.
\label{E:how1}
\end{equation}
It follows from (\ref{E:how1}) and (\ref{E:diff2}) that
\begin{equation}
\gamma=\frac{2\delta\left(y_1\cos\left(\frac{\delta}{2}\right)-y_0\right)}
{\sin\left(\frac{\delta}{2}\right)}.
\label{E:how3}
\end{equation}
Moreover, (\ref{E:how1}) and (\ref{E:diff1}) imply that $\delta^{*}$
is a solution to the following equation:
\begin{align}
&\frac{4y_0^2\delta^2\left(\delta+\sin(\delta)\right)
+\gamma^2\left(\delta-\sin(\delta)\right)
+4\delta^2\gamma y_0\sin\left(\frac{\delta}{2}\right)}
{8\delta^2\cos^2\left(\frac{\delta}{2}\right)} \nonumber \\
&=x_1-x_0.
\label{E:how4}
\end{align}

The second component $\gamma^{*}$ of the critical point $(\delta^{*},\gamma^{*})$ can be found by plugging the solution $\delta^{*}$ to the equation
(\ref{E:how4}) into (\ref{E:how3}). This gives
\begin{equation}
\gamma^{*}=\frac{2\delta^{*}\left(y_1\cos\left(\frac{\delta^{*}}{2}\right)-y_0\right)}
{\sin\left(\frac{\delta^{*}}{2}\right)}.
\label{E:howw}
\end{equation}

Our next goal is to simplify the equation in (\ref{E:how4}) by taking into account (\ref{E:how4}) and (\ref{E:howw}). Tedious, but rather straightforward calculations show that
(\ref{E:how4}) can be rewritten in the following form:
\begin{equation}
\tilde{f}(\delta)=\tilde{f}(\delta,y_0,y_1)=x_1-x_0,
\label{E:can1}
\end{equation}
where
\begin{equation}
\tilde{f}(\delta)=\frac{1}{2}\tilde{A}(\delta)\tilde{B}(\delta)
\label{E:cand}
\end{equation}
with
\begin{equation}
\tilde{A}(\delta)=\tilde{A}(\delta,y_0,y_1)=\sin^{-2}\left(\frac{\delta}{2}\right)
\label{E:can2}
\end{equation}
and
\begin{align}
&\tilde{B}(\delta)=\tilde{B}(\delta, y_0,y_1)=\left(y_1^2+y_0^2\right)
\left(\delta-\sin(\delta)\right) \nonumber \\
&\quad-2y_1y_0\left(\delta\cos\left(\frac{\delta}{2}\right)
-2\sin\left(\frac{\delta}{2}\right)\right).
\label{E:can3}
\end{align}
It is assumed in (\ref{E:can1}) that $-\pi\le\delta\le\pi$. The value of the function $\tilde{f}$ at $\delta=0$ is
computed as follows:
$\displaystyle{\tilde{f}(0)=\frac{1}{2}\lim_{\delta\rightarrow 0}\tilde{A}(\delta)\tilde{B}(\delta)=0}$.
\begin{remark}\label{R:innt} \rm
It follows from (\ref{E:cand})\,-\,(\ref{E:can3}) that the function $f$ is defined on the interval $(-2\pi,2\pi)$. This fact will be used below.
\end{remark}
\begin{remark}\label{R:simil} \rm
The properties of the functions $\tilde{f}$, $\tilde{A}$, and $\tilde{B}$ are similar to those of the functions $f$, $A$, and $B$ defined by
(\ref{E:cand0}), (\ref{E:can20}), and (\ref{E:can30}), respectively. The previous statement follows from
the equalities
$\tilde{f}(\delta,y_0,y_1)=f(\delta,y_0^2,y_1^2)$, $\tilde{A}(\delta,y_0,y_1)=A(\delta,y_0^2,y_1^2)$,
$\tilde{B}(\delta,y_0,y_1)=B(\delta,y_0^2,y_1^2)$,
and
$\delta^{*}((x_0,y_0),(x_1,y_1))=\hat{\delta}((x_0,y_0),(x_1,y_1))$.
\end{remark}

Let us fix $x_0$ and $y_0$. It follows from Lemma 2.3.9 in \cite{DZ} that the Legendre-Fenchel transform $\Lambda^{*}$ of
$\Lambda$ is a good rate function.
Explicit formulas for the function $\Lambda$ were found in
Section \ref{S:lcgf}. In the present section, we compute the function $\Lambda^{*}$. A simple analysis
of formula (\ref{E:cp1}) defining the function $\Lambda^{*}$ shows that to compute the supremum in  (\ref{E:cp1}) one has to take into account
the input of the critical point
$(\delta^{*},\gamma^{*})$, the boundary of the strip where the moment generating function is finite, and the boundary at infinity. Since $\Lambda(\delta,\gamma)\rightarrow\infty$ as $\gamma\rightarrow\infty$ or $\gamma\rightarrow-\infty$,
the input of the boundary at infinity can be ignored. Using formulas (\ref{E:insert1}) and (\ref{E:insert2}),
we see that the the input of the exceptional points $P_1$ and $P_2$, more precisely, of sequences converging to those points, is given by the following expression:
\begin{equation}
R_1=\pi(|x_0-x_1|-2y_1y_0).
\label{E:input}
\end{equation}
Note that the number in (\ref{E:input}) is positive if and only if
$|x_0-x_1|> 2y_1y_0$.

Next, suppose $\delta^{*}\notin(-\pi,\pi)$. This means that
\begin{equation}
\frac{1}{2}\left[\pi\left(y_1^2+y_0^2\right)+4y_1y_0\right]\le|x_1-x_0|.
\label{E:crr}
\end{equation}
In this case, there is no critical point inside the fundamental strip, and hence $\Lambda^{*}$ is given by
\begin{equation}
\Lambda^{*}(x_0,y_0,x_1,y_1)=\pi(|x_1-x_0|-2y_1y_0)=R_1.
\label{E:vdrug3}
\end{equation}

On the other hand, if $\delta^{*}\in(-\pi,\pi)$ (this happens if the opposite inequality to the inequality in
(\ref{E:crr}) holds), then the input that the critical point $(\delta^{*},\gamma^{*})$ brings to the computation of the supremum in the formula for $\Lambda^{*}$ 
is given by the following expression:
\begin{align}
R_2&=(x_1-x_0)\delta^{*}+y_1\gamma^{*} \nonumber \\
&\quad-\frac{\left(4\left(\delta^{*}\right)^2y_0^2+\left(\gamma^{*}\right)^2\right)\sin\left(\frac{\delta^{*}}{2}\right)
+4\gamma^{*}\delta^{*}y_0}{4\delta^{*}\cos\left(\frac{\delta^{*}}{2}\right)}.
\label{E:in1}
\end{align}
Replacing $x_1-x_0$ in formula (\ref{E:in1}) by the expression on the left-hand side of
formula (\ref{E:can1}) and making simplifications, we obtain
\begin{equation}
R_2=\frac{(\delta^{*})^2}{2\sin^2\left(\frac{\delta^{*}}{2}\right)}\left[y_1^2+y_0^2-2y_1y_0\cos\left(\frac{\delta^{*}}{2}\right)\right].
\label{E:finst1}
\end{equation}
Therefore the condition $|x_1-x_0|\le 2y_1y_0$ implies the equality
\begin{equation}
\Lambda^{*}(x_0,y_0,x_1,y_1)=R_2,
\label{E:doom1}
\end{equation}
and the condition $2y_1y_0<|x_1-x_0|<\frac{1}{2}\left[\pi\left(y_1^2+y_0^2\right)+4y_1y_0\right]$ gives
\begin{equation}
\Lambda^{*}(x_0,y_0,x_1,y_1)
=\rm\max\left\{R_1,R_2\right\}.
\label{E:doom2}
\end{equation}

Our next goal is to compare all the inputs discussed above. We will next show that $R_2$ always dominates $R_1$.
\begin{lemma}\label{L:domi}
For all $(x_0,y_0)\in\mathbb{R}^2$ and $(x_1,y_1)\in\mathbb{R}^2$, the following inequality
holds: $R_1\le R_2$.
In addition, if
\begin{equation}
\frac{1}{2}\left[\pi\left(y_1^2+y_0^2\right)+4y_1y_0\right]<|x_1-x_0|,
\label{E:crrr}
\end{equation}
then $R_1< R_2$.
\end{lemma}

\it Proof. \rm
Taking into accound the definition of $R_1$ and $R_2$ (see (\ref{E:input}) and (\ref{E:in1}), respectively),
replacing the expression $x_1-x_0$ in formula (\ref{E:in1}) by the expression on the left-hand side of
formula (\ref{E:can1}), and simplifying, we see that the inequality $R_1\le R_2$
can de derived from the inequality
\begin{align}
&\left(y_1^2+y_0^2\right)\left[(\delta^{*})^2+\pi\sin\left(\delta^{*}\right)\right] \nonumber \\
&\quad+y_1y_0\left[2\pi\delta^{*}\cos\left(\frac{\delta^{*}}{2}\right)
+4\pi\sin^2\left(\frac{\delta^{*}}{2}\right)\right] \nonumber \\
&\ge\left(y_1^2+y_0^2\right)\pi\delta^{*}+y_1y_0\left[2(\delta^{*})^2\cos\left(\frac{\delta^{*}}{2}\right)
+4\pi\sin\left(\frac{\delta^{*}}{2}\right)\right].
\label{E:glavnoe}
\end{align}
It is easy to see that with no loss of generality we may assume that $\delta^{*}> 0$.
We will prove that
\begin{equation}
(\delta^{*})^2+\pi\sin\left(\delta^{*}\right)\ge\pi\delta^{*}
\label{E:esst1}
\end{equation}
and
\begin{align}
&\pi\delta^{*}\cos\left(\frac{\delta^{*}}{2}\right)
+2\pi\sin^2\left(\frac{\delta^{*}}{2}\right) \nonumber \\
&\quad\ge(\delta^{*})^2\cos\left(\frac{\delta^{*}}{2}\right)
+2\pi\sin\left(\frac{\delta^{*}}{2}\right)
\label{E:esst2}
\end{align}
for all $0<\delta^{*}\le\pi$. Moreover, it will be shown that the condition $\pi<\delta^{*}<2\pi$
implies strict inequalities in (\ref{E:esst1}) and (\ref{E:esst2}). Note that condition (\ref{E:crrr})
is equivalent to the condition $\pi<\delta^{*}<2\pi$.
It is clear that Lemma \ref{L:domi} follows
from the inequalities formulated above.

Suppose first that $0<\delta^{*}\le\pi$.
We will next establish (\ref{E:esst1}). The fact that the inequality in (\ref{E:esst2}) is equivalent to the inequality
\begin{equation}
\delta^{*}(\pi-\delta^{*})\cos\left(\frac{\delta^{*}}{2}\right)\ge 2\pi\sin\left(\frac{\delta^{*}}{2}\right)
\left(1-\sin\left(\frac{\delta^{*}}{2}\right)\right)
\label{E:esst3}
\end{equation}
will be used in the proof.

Let us assume that (\ref{E:esst1}) holds for all $0<\delta^{*}<\frac{\pi}{2}$, and let
$\frac{\pi}{2}<\delta^{*}<\pi$. Then $\tilde{\delta}=\pi-\delta^{*}$ satisfies (\ref{E:esst1}),
and it is easy to see that $\delta^{*}$ also satisfies
(\ref{E:esst1}). It follows that it suffices to assume $0<\delta^{*}<\frac{\pi}{2}$.

Using the Taylor series, we see that for $0<\delta^{*}<\frac{\pi}{2}$,
$$
\frac{\sin(\delta^{*})}{\delta^{*}}\ge 1-\frac{(\delta^{*})^2}{6}.
$$
Hence (\ref{E:esst1}) can be derived from the inequality
$$
\pi\left(1-\frac{(\delta^{*})^2}{6}\right)\ge\pi-\delta^{*}.
$$
The previous inequality is equivalent to $\pi\delta^{*}\le 6$,
which is of course correct. This establishes (\ref{E:esst1}).

Our next goal is to prove (\ref{E:esst3}). Let us first assume $0<\delta^{*}<\frac{\pi}{2}$.
Using the Taylor series, we obtain
\begin{equation}
\cos\left(\frac{\delta^{*}}{2}\right)\ge 1-\frac{(\delta^{*})^2}{8}\quad\mbox{and}\quad\sin\left(\frac{\delta^{*}}{2}\right)
\ge\frac{\delta^{*}}{2}-\frac{(\delta^{*})^3}{48}.
\label{E:taylor}
\end{equation}
Therefore, (\ref{E:esst3}) can be derived from the following inequality:
$$
\delta^{*}\left(\pi-\delta^{*}\right)\left(1-\frac{(\delta^{*})^2}{8}\right)\ge\pi\delta^{*}\left(1-\frac{\delta^{*}}{2}
+\frac{(\delta^{*})^3}{48}\right).
$$
The previous inequality is equivalent to
\begin{equation}
6(\delta^{*})^2+24\pi\ge 48+6\pi\delta^{*}+\pi(\delta^{*})^2.
\label{E:qr1}
\end{equation}
It is not hard to see that (\ref{E:qr1}) can be rewritten as follows:
\begin{equation}
(6-\pi)\left[\left(\delta^{*}-\frac{3\pi}{6-\pi}\right)^2
+\frac{24\pi-48}{6-\pi}-\frac{9\pi^2}{(6-\pi)^2}\right]\ge 0.
\label{E:qr2}
\end{equation}
Since $\delta^{*}<\frac{\pi}{2}$, (\ref{E:qr2}) follows from
\begin{equation}
(6-\pi)\left[\left(\frac{3\pi}{6-\pi}-\frac{\pi}{2}\right)^2
+\frac{24\pi-48}{6-\pi}-\frac{9\pi^2}{(6-\pi)^2}\right]\ge 0.
\label{E:qr3}
\end{equation}
Now, it is clear that (\ref{E:esst3}) holds for all $0<\delta^{*}<\frac{\pi}{2}$,
since (\ref{E:qr3}) can be easily checked.

We will next show that (\ref{E:esst3}) also holds under the condition
$\frac{\pi}{2}\le\delta^{*}<\pi$.
The following estimate will be needed in the sequel. For all $\tilde{\delta}$ with $0<\tilde{\delta}<\frac{\pi}{2}$,
\begin{equation}
\tilde{\delta}(\pi-\tilde{\delta})\sin\left(\frac{\tilde{\delta}}{2}\right)\ge 2\pi\cos\left(\frac{\tilde{\delta}}{2}\right)
\left(1-\cos\left(\frac{\tilde{\delta}}{2}\right)\right).
\label{E:esst4}
\end{equation}
Dividing the both sides of (\ref{E:esst4}) by $\cos\left(\frac{\tilde{\delta}}{2}\right)$
and using the inequality $\tan x\ge x$ for
$0< x<\frac{\pi}{2}$, we see that (\ref{E:esst4}) can be derived from the inequality
\begin{equation}
\tilde{\delta}^2(\pi-\tilde{\delta})\ge 4\pi\left(1-\cos\left(\frac{\tilde{\delta}}{2}\right)\right).
\label{E:esst5}
\end{equation}
Next, using the first inequality in (\ref{E:taylor}), we see that (\ref{E:esst5})
follows from the inequality
\begin{equation}
(\pi-\tilde{\delta})\ge\frac{\pi}{2},
\label{E:eqeq}
\end{equation}
which clearly holds because
$0<\tilde{\delta}<\frac{\pi}{2}$.

Suppose that $\frac{\pi}{2}<\delta^{*}<\pi$ and denote $\tilde{\delta}=\pi-\delta^{*}$. Then $0<\tilde{\delta}<\frac{\pi}{2}$, and hence inequality
(\ref{E:eqeq}) holds for $\tilde{\delta}$. It is not hard to see that (\ref{E:eqeq}) for
$\tilde{\delta}$ is equivalent to (\ref{E:esst3})
for $\delta^{*}$. This completes the proof of estimate (\ref{E:esst2}).

It follows from (\ref{E:esst1}) and (\ref{E:esst2}) that esimate (\ref{E:glavnoe}) holds. It has already been mentioned that (\ref{E:glavnoe})
implies the inequality $R_1\le R_2$. Therefore, part of Lemma \ref{L:domi} in the
case where $-\pi\le\delta^{*}\le\pi$ is valid.

Now let $\pi<\delta^{*}<2\pi$. We will first establish that the strict inequality in (\ref{E:esst1})
holds. It is clear that the function $\rho_1(u)$ on the left-hand side of
(\ref{E:esst1}) and the function $\rho_2(u)$ on the right-hand side equal $\pi^2$ at $u=\pi$.
Moreover, $\rho_1^{\prime}(u)=2u+\pi\cos u$ and $\rho_2^{\prime}(u)=\pi$. It is easy to see that
$\rho_1^{\prime}(u)>\rho_2^{\prime}(u)$ for all $\pi< u<2\pi$. It follows that the strict inequality in (\ref{E:esst1}) holds when $\pi<\delta^{*}<2\pi$. The proof of the strict inequality in (\ref{E:esst2})
under the same restriction is similar. Here the functions $\rho_1$ and $\rho_2$ equal $2\pi$
at $u=\pi$. Moreover
$$
\rho_1^{\prime}(u)=\pi\cos\frac{u}{2}-\frac{\pi}{2}u\sin\frac{u}{2}+2\pi\sin\frac{u}{2}\cos\frac{u}{2}
$$
and
$$
\rho_2^{\prime}(u)=\pi\cos\frac{u}{2}-\frac{1}{2}u^2\sin\frac{u}{2}+2u\cos\frac{u}{2}.
$$
It is not hard to see that $\rho_1^{\prime}(u)>\rho_2^{\prime}(u)$ for all $\pi< u<2\pi$. This implies
the strict inequality in (\ref{E:esst2}) in the case where $\pi<\delta^{*}<2\pi$.

The proof of Lemma \ref{L:domi} is thus completed.
\begin{theorem}\label{T:glavnaja}
Under the condition
\begin{equation}
|x_1-x_0|\le\frac{1}{2}\left[\pi\left(y_1^2+y_0^2\right)+4y_1y_0\right],
\label{E:coon1}
\end{equation}
the Legendre-Fenchel transform $\Lambda^{*}$ of the limiting cumulant generating function $\Lambda$ in the Grushin model is given by the following formula:
\begin{equation}
\Lambda^{*}(x_0,y_0,x_1,y_1)=\frac{(\delta^{*})^2}{2\sin^2\left(\frac{\delta^{*}}{2}\right)}\left(y_1^2+y_0^2-2y_1y_0\cos\left(\frac{\delta^{*}}{2}\right)\right).
\label{E:fins1}
\end{equation}
On the other hand, if
\begin{equation}
\frac{1}{2}\left[\pi\left(y_1^2+y_0^2\right)+4y_1y_0\right]<|x_1-x_0|,
\label{E:coon2}
\end{equation}
then we have
\begin{equation}
\Lambda^{*}(x_0,y_0,x_1,y_1)=\pi(|x_1-x_0|-2y_1y_0).
\label{E:fin3}
\end{equation}
\end{theorem}
\begin{remark} \rm
Recall that $\delta^{*}$ in formula (\ref{E:fins1}) is the unique solution to the equation
\begin{align*}
&\frac{1}{2}\csc^2\left(\frac{\delta}{2}\right)
\left[\left(y_1^2+y_0^2\right)
\left(\delta-\sin(\delta)\right)
-2y_1y_0\left(\delta\cos\left(\frac{\delta}{2}\right)-2\sin\left(\frac{\delta}{2}\right)\right)\right] \\
&=x_1-x_0.
\end{align*}
The solution to this equation satisfies $-\pi<\delta^{*}<\pi$ if condition (\ref{E:coon1}) holds.
\end{remark}

\it Proof of Theorem \ref{T:glavnaja}. \rm Formula (\ref{E:fin3}) in Theorem \ref{T:glavnaja} has already been established (see (\ref{E:vdrug3})).
Formula (\ref{E:fins1}) can be derived from
(\ref{E:finst1}), (\ref{E:doom1}), (\ref{E:doom2}), and Lemma \ref{L:domi}.
\begin{remark}\label{R:vert} \rm
Let us consider a special case where $x_1=x_0$. Then $\delta^{*}=0$ and therefore (\ref{E:fins1}) gives
\begin{equation}
\Lambda^{*}(x_0,y_0,x_1,y_1)=2(y_1-y_0)^2.
\label{E:horf}
\end{equation}
\end{remark}
\section{A partial large deviation principle for the Grushin model}\label{S:pldp}
Le us recall that in the present paper we denoted by $\Lambda$ and $\Lambda^{*}$ the limiting cumulant generating function in the Grushin model and the Legendre-Fenchel transform of $\Lambda$, respectively. We will prove below that for any initial point $A=(x_0,y_0)\in\mathbb{R}^2$ the large deviation principle holds for the Grushin model in a certain open subset $M$ of the plane $\mathbb{R}^2$.  
The set $M$ consists of all the points which are $\delta$-close to $A$. In the formulation of the next theorem, the symbols $B^{\circ}$ and $\overline{B}$ stand for the interior and the closure of the set $B$,
respectively.
\begin{theorem}\label{T:sum}
Let $x_0\in\mathbb{R}$ and $y_0\in\mathbb{R}$ be fixed, and
consider the open set in $\mathbb{R}^2$ defined by
\begin{equation}
M=\left\{(x_1,y_1)\in\mathbb{R}^2:|x_1-x_0|<\frac{1}{2}\left[\pi\left(y_1^2+y_0^2\right)
+4y_1y_0\right]\right\}.
\label{E:MM}
\end{equation}
Then the large deviation principle with the rate function $\Lambda^{*}$ holds on the set $M$. More precisely, for any Borel subset $B$ of $M$,
\begin{align}
&-\inf_{(x_1,y_1)\in B^{\circ}}\Lambda^{*}(x_0,y_0,x_1,y_1)
\le\liminf_{t\rightarrow 0}\left[t
\log P^G_t(x_0,y_0,B)\right] \nonumber \\
&\le\limsup_{t\rightarrow 0}\left[t\log P^G_t(x_0,y_0,B)\right]
\le-\inf_{(x_1,y_1)\in\overline{B}}\Lambda^{*}(x_0,y_0,x_1,y_1).
\label{E:ldd1}
\end{align}
In addition, if the set $B$ is such that $\overline{B^{\circ}}=\overline{B}$, then
\begin{equation}
\lim_{t\rightarrow 0}\left[t\log P^G_t(x_0,y_0,B)\right]
=-\inf_{(x_1,y_1)\in B}\Lambda^{*}(x_0,y_0,x_1,y_1).
\label{E:ldpe}
\end{equation}
\end{theorem}

\it Proof. \rm
We have shown in Section \ref{S:how}
that for all $(x_1,y_1)\in M$, the following equality holds:
$\nabla\Lambda(\delta^{*},\gamma^{*})=(x_1,y_1)$.
Therefore, Lemma 2.3.9 in \cite{DZ} implies that any point $(x_1,y_1)\in M$ is an exposed point of $\Lambda^{*}$ with $(\delta^{*},\gamma^{*})$
being the exposing hyperplane
for $(x_1,y_1)$ (see \cite{DZ} for the definition of exposed points and exposing hyperplanes). In other words, the set $M$ consists entirely of exposed points. Now it is not hard to see that all the conditions in the G\"{a}rtner-Ellis theorem (see Theorem 2.3.6 in \cite{DZ}) hold. Applying this theorem and taking into account the continuity of the function $\Lambda^{*}$, we establish Theorem \ref{T:sum}.
\section{The distance formula}\label{S:df}
Our goal in this section is to complete the proof of Theorem \ref{T:Hd2}. Since we have already proved this theorem in the far $\delta$-regime, it remains to establish it for the pairs of points
$(x_0,v_0)\in{\cal H}^{\circ}$ and $(x_1,v_1)\in{\cal H}^{\circ}$ in the close $\delta$-regime.

We will prove the following assetrion:
\begin{theorem}\label{T:esst}
If condition (\ref{E:coon1}) holds, then
\begin{equation}
\frac{d_G^2((x_0,y_0),(x_1,y_1))}{2}=\Lambda^{*}(x_0,y_0,x_1,y_1),
\label{E:costy1}
\end{equation}
while if condition (\ref{E:coon2}) is valid, then
\begin{equation}
\frac{d_G^2((x_0,y_0),(x_1,y_1))}{2}>\Lambda^{*}(x_0,y_0,x_1,y_1).
\label{E:costy2}
\end{equation}
\end{theorem}
\begin{remark}\label{R:folllows} \rm
It is clear that Theorem \ref{T:Hd2} in the close $\delta$-regime follows from (\ref{E:GruHes}), (\ref{E:fins1}) and (\ref{E:costy1}). Inequality (\ref{E:costy2}) is interesting because it is often expected that for the distances associated with stochastic models, the function $\frac{1}{2}d^2$
coincides with the Legendre-Fenchel transform $\Lambda^{*}$ of the limiting cumulant generating function 
$\Lambda$. Inequality
(\ref{E:costy2}) shows that this is not the case for the Grushin model when the points are in the far
$\delta$-regime.
\end{remark}

\it Proof of Theorem \ref{T:esst}. \rm Since the Grushin operator (\ref{E:Gruo}) is hypoelliptic, it follows from the results obtained by L\'{e}andre in \cite{Le1}, \cite{Le2} (see also \cite{Le3}, \cite{Le4})
that Varadhan's equality, that is, the equality
\begin{equation}
\lim_{t\rightarrow 0}[t\log p_t^G((x_0,y_0),(x_1,y_1))]=-\frac{d_G^2((x_0,y_0),(x_1,y_1))}{2}
\label{E:varvar}
\end{equation}
holds uniformly on compact subsets of $\mathbb{R}^2$ (Varadhan's results can be found in \cite{V1,V2}).

Fix a point $(x_0,y_0)\in\mathbb{R}^2$
and consider all the points $(x_1,y_1)\in\mathbb{R}^2$ such that condition
(\ref{E:coon1}) holds. Let $B=B_{\varepsilon}(x_1,y_1)$,
where $B_{\varepsilon}(x_1,y_1)$ is the disk of radius $\varepsilon$
in $\mathbb{R}^2$ centered at $(x_1,y_1)$ and such that
$$
B_{\varepsilon}(x_1,y_1)\subset M
$$
(recall that $M$ is defined in (\ref{E:MM})). It follows from
(\ref{E:ldpe}) that
\begin{align}
&\lim_{t\rightarrow 0}\left[t\log P_t\left(x_0,y_0,B_{\varepsilon}(x_1,y_1)\right)\right]
\nonumber \\
&=-\inf_{(x,y)\in B_{\varepsilon}(x_1,y_1)}\Lambda^{*}(x_0,y_0,x,y)
\label{E:ldpr}
\end{align}
for all small enough $\varepsilon> 0$. Using the mean value theorem for integrals, we can prove that
\begin{align}
&\left|t\log P_t^G\left(x_0,y_0,B_{\varepsilon}(x_1,y_1)\right)
+\frac{d_G^2((x_0,y_0),(x_1,y_1))}{2}\right|
\le\left|t\log(\pi\varepsilon^2)\right| \nonumber \\
&+\sup_{(x,y)\in B_{\varepsilon}(x_1,y_1)}\left|t\log p_t\left(x_0,x_1,x,y\right)
+\frac{d_G^2\left((x_0,y_0),\left(x,y\right)\right)}{2}\right| \nonumber \\
&+\sup_{(x,y)\in B_{\varepsilon}(x_1,y_1)}\left|\frac{d_G^2((x_0,y_0),(x_1,y_1))}{2}-\frac{d_G^2\left((x_0,y_0),\left(x,y\right)\right)}{2}\right|.
\label{E:disf}
\end{align}
It follows from (\ref{E:disf}) and (\ref{E:varvar}) that
\begin{align}
&\limsup_{t\rightarrow 0}\left|t\log P_t^G\left(x_0,y_0,B_{\varepsilon}(x_1,y_1)\right)+\frac{d_G^2((x_0,y_0),(x_1,y_1))}{2}\right|
\nonumber \\
&\le\sup_{(x,y)\in B_{\varepsilon}(x_1,y_1)}\left|\frac{d_G^2((x_0,y_0),(x_1,y_1))}{2}
-\frac{d_G^2\left((x_0,y_0),\left(x,y\right)\right)}{2}\right|.
\label{E:vl1}
\end{align}
Taking the limit as $\varepsilon\rightarrow 0$ in (\ref{E:vl1}) and using the continuity of the function $d^2$ (the continuity follows from (\ref{E:twos2})), we get
\begin{align}
&\lim_{\varepsilon\rightarrow 0}\limsup_{t\rightarrow 0}\left|t\log P_t^G\left(x_0,y_0,B_{\varepsilon}(x_1,y_1)\right)+\frac{d_G^2((x_0,y_0),(x_1,y_1))}{2}\right|
\nonumber \\
&=0.
\label{E:vl2}
\end{align}
It is not hard to see that
\begin{align}
&\left|\frac{d_G^2((x_0,y_0),(x_1,y_1))}{2}-\Lambda^{*}(x_0,y_0,x_1,y_1)\right| \nonumber \\
&\le\left|t\log P_t^G\left(x_0,y_0,B_{\varepsilon}(x_1,y_1)\right)+\frac{d_G^2((x_0,y_0),(x_1,y_1))}{2}\right|
\nonumber \\
&\quad+\left|t\log P_t^G\left(x_0,y_0,B_{\varepsilon}(x_1,y_1)\right)+\inf_{(x,y)\in B_{\varepsilon}(x_1,y_1)}\Lambda^{*}\left(x_0,y_0,x,y\right)\right|
\nonumber \\
&\quad+\left|\inf_{(x,y)\in B_{\varepsilon}(x_1,y_1)}\Lambda^{*}\left(x_0,y_0,x,y\right)-\Lambda^{*}(x_0,y_0,x_1,y_1)\right|.
\label{E:vl3}
\end{align}
Taking the limit as $\varepsilon\rightarrow 0$ in (\ref{E:vl3}) and using (\ref{E:ldpr}), (\ref{E:vl2}), and the continuity of the function $\Lambda^{*}$,
we obtain formula (\ref{E:costy1})
for all pairs of points $(x_0,y_0)$, $(x_1,y_1)$ satisfying condition (\ref{E:coon1}). In addition, it is not hard to see that formula (\ref{E:costy2}) for all pairs of points $(x_0,y_0)$, $(x_1,y_1)$ satisfying condition (\ref{E:coon2}) follows from (\ref{E:fin3}), (\ref{E:input}), (\ref{E:finst1}), the second part of
Lemma \ref{L:domi}, and from Theorem \ref{T:Hd2} in the far $\delta$-regime. 
Note that we have already established Theorem \ref{T:Hd2} for points which are $\delta$-far.

This completes the proof of Theorem \ref{T:esst}.

\end{document}